\documentclass[fleqn,usenatbib]{mnras}

\usepackage{newtxtext,newtxmath}
\usepackage{lscape}

\usepackage[T1]{fontenc}
\usepackage{xcolor}

\DeclareRobustCommand{\VAN}[3]{#2}
\let\VANthebibliography\thebibliography
\def\thebibliography{\DeclareRobustCommand{\VAN}[3]{##3}\VANthebibliography}

\usepackage{graphicx}	
\usepackage{float}

\newcommand{\Gyr}             {\rm{Gyr}}
\newcommand{\ofe}             {[\rm O/Fe]}
\newcommand{\afe}             {[\upalpha/{\rm Fe}]}
\newcommand{\feh}             {[\rm Fe/H]}

\newcommand{\hMsun}            {h^{-1}\rm{M}_\odot}

\newbox\grsign \setbox\grsign=\hbox{$>$} \newdimen\grdimen \grdimen=\ht\grsign
\newbox\simlessbox \newbox\simgreatbox
\setbox\simgreatbox=\hbox{\raise.5ex\hbox{$>$}\llap
     {\lower.5ex\hbox{$\sim$}}}\ht1=\grdimen\dp1=0pt
\setbox\simlessbox=\hbox{\raise.5ex\hbox{$<$}\llap
     {\lower.5ex\hbox{$\sim$}}}\ht2=\grdimen\dp2=0pt
\def\simgreater{\mathrel{\copy\simgreatbox}}

\newbox\simppropto
\setbox\simppropto=\hbox{\raise.5ex\hbox{$\sim$}\llap
     {\lower.5ex\hbox{$\propto$}}}\ht2=\grdimen\dp2=0pt

\title[The origin of $\upalpha$~knees]{Realistic simulated galaxies form [$\upalpha$/Fe]-[Fe/H] knees due to a sustained decline in their star formation rates}

\author[]{\noindent Andrew C.  Mason,$^{1}$\thanks{E-mail: A.C.Mason@2015.ljmu.ac.uk} Robert A. Crain,$^{1}$ Ricardo P. Schiavon,$^{1}$ David H. Weinberg,$^{2}$ Joel Pfeffer,$^{3}$\newauthor Joop Schaye,$^{4}$ Matthieu Schaller,$^{4,5}$ and Tom Theuns$^{6,7}$
\\
$^{1}$Astrophysics Research Institute, Liverpool John Moores University, 146 Brownlow Hill, Liverpool, L3 5RF\\
$^{2}$The Department of Astronomy and Centre of Cosmology and AstroParticle Physics, The Ohio State University, Columbus, OH 43210, USA\\
$^{3}$International Centre for Radio Astronomy Research (ICRAR), M468, University of Western Australia, 35 Stirling Hwy, Crawley, WA 6009, Australia\\
$^{4}$Leiden Observatory, Leiden University, PO Box 9513 2300 RA Leiden, the Netherlands\\
$^{5}$Lorentz Institute for Theoretical Physics, Leiden Univeristy, PO box 9506, 2300 RA Leiden, the Netherlands\\
$^6$ Institute for Computational Cosmology, Durham University, South Road,Durham DH1 3LE, UK\\
$^{7}$Department of Physics, Durham University, South Road, Durham DH1 3LE, UK
}   
\date{Accepted XXX.  Received YYY; in original form ZZZ}

\pubyear{2023}

\begin{document}
\label{firstpage}
\pagerange{\pageref{firstpage}--\pageref{lastpage}}
\maketitle

\begin{abstract}
\noindent We examine the stellar [$\upalpha$/Fe]-[Fe/H] distribution of $\simeq1000$ present-day galaxies in a high-resolution EAGLE simulation. A slight majority of galaxies exhibit the canonical distribution, characterised by a sequence of low-metallicity stars with high [$\upalpha$/Fe] that transitions at a `knee' to a sequence of declining [$\upalpha$/Fe] with increasing metallicity. 
This population yields a knee metallicity - galaxy-mass relation similar to that observed in Local Group galaxies, both in slope and scatter. However, many simulated galaxies lack a knee or exhibit more complicated distributions. Knees are found only in galaxies with star formation histories (SFHs) featuring a sustained decline from an early peak ($t\simeq7~{\rm Gyr}$), which enables enrichment by Type Ia supernovae to dominate that due to Type II supernovae (SN~II), reducing [$\upalpha$/Fe] in the interstellar gas. 
The simulation thus indicates that, contrary to the common interpretation implied by analytic galactic chemical evolution (GCE) models, knee formation is not a consequence of the onset of enrichment by SN~Ia. 
We use the SFH of a simulated galaxy exhibiting a knee as input to the \textsc{vice} GCE model, finding it yields an $\alpha$-rich plateau enriched only by SN~II, but the plateau comprises little stellar mass and the galaxy forms few metal-poor ([Fe/H]$\lesssim$--1) stars. This follows from the short, constant gas consumption timescale typically assumed by GCEs, which implies the presence of a readily-enriched, low-mass gas reservoir. When an initially longer, evolving consumption timescale is adopted, \textsc{vice} reproduces the simulated galaxy's track through the [$\upalpha$/Fe]-[Fe/H] plane and its metallicity distribution function.
\end{abstract}

\begin{keywords}
methods: numerical -- methods: analytical -- galaxies:abundances -- galaxies: stellar content -- galaxies:Local Group -- galaxies:evolution 
\end{keywords}



\section{Introduction}\label{seci_intro}
Galactic archaeology has undergone a renaissance in recent years with the emergence of a number of ambitious stellar surveys.  Precision element abundances and phase space properties of millions of stars have now been measured and interpreted within the broader context of theoretical predictions from galaxy formation theory to help elucidate the formation history of the Galaxy. Such efforts include RAVE \citep{RAVE}, SEGUE \citep{SEGUE}, Gaia-ESO \citep{gilmore-geso,randich-geso},  GALAH \citep{GALAH,GALAH2}, APOGEE \citep{apogee}, H3 \citep{H3}, and the Gaia mission \citep{gaia}, as well as surveys forthcoming in the near future \citep[e.g.,][]{4most,MOONS2,weave}.

It has been suggested that the Milky Way exhibits properties somewhat atypical of similarly massive disc galaxies \citep[see, e.g.,][]{mackereth-bimodality,ripisc-MW-symposium}, implying that study of the Galaxy alone cannot offer a comprehensive picture of galaxy formation and evolution. Besides the Galaxy, the Local Group also hosts a population of dwarf galaxies ($M_\star\lesssim10^{10}~\rm{M}_{\odot}$) which comprise both `fossil' and young stellar populations. Dwarfs are the most abundant galaxies in the Universe (e.g.  \citealp{li-gsmf}; \citealp{baldry_gama}) and yet detailed observations of their stellar populations are comparatively sparse compared to the Milky Way.  
The Local Group is thought to host at least $\simeq$150 (e.g.  \citealp{drlica-lgds, newton-lgds}) dwarf galaxies, which are observed to exhibit diverse star formation histories, even at fixed $M_\star$ and even within the same morphological class \citep[see, e.g.,][]{dwarfs-sfh-1-weisz,dwarfs-sfh-2-weisz,dwarfs-sfh-3-gallart,dwarfs-sfh-4-skillman}.  The Local Group in totality thus represents a rich parameter space and understanding how the panoply of its formation histories maps onto the so-called canonical chemical planes in the literature remains an open question.

For nearby galaxies, the resolvable distribution of individual stars in the [$\upalpha$/Fe]-[Fe/H] plane (hereafter $\upalpha$-Fe plane, and sometimes referred to as the Tinsley-Wallerstein diagram following \citealp{tinsley-chemistry} and \citealp{wallerstein_62}) encodes a wealth of information about the galaxy's formation history.  
The distribution exhibited by galaxies is affected by, amongst other things, the timescales of production of the $\upalpha$-elements (i.e. O, Mg, Si, Ca, Ti) and Fe.  The former are synthesised by the progenitors of core-collapse supernovae \citep[hereafter SN~II, e.g.][]{portinari-yields, kobayashi_06} whereas the latter is synthesised by both SN~II and Type Ia supernovae \citep[hereafter SN~Ia, e.g.][]{woosley94, iwamoto_1a, seitenzahl13, gronow21a, gronow21b}. 
The two classes of SN explode and return their nucleosynthetic products to the interstellar medium (ISM) on markedly different timescales, fostering a time dependence of the characteristic ratio between the abundances of $\upalpha$-elements and iron, [$\upalpha$/Fe], in the ejecta produced by a simple stellar population.

The early phase of chemical evolution stemming from a nascent stellar population is dominated by SN~II due to the short lifetimes of their progenitors ($\tau\simeq3-30~\rm{Myr}$), whose masses range between $M_{\star}\simeq8$ and $100~\rm{M}_{\odot}$ \citep[e.g.,][]{portinari-yields}.  
The hydrostatic $\upalpha$-elements O and Mg are produced during the hydrostatic burning stage over the progenitor's lifetime while the explosive $\upalpha$-elements (Si, Ti, and Ca) are produced in the SN explosion.

SN~Ia originate from white dwarf (WD) stars in binaries that experience mass transfer from their companion stars, exceeding the Chandrasekhar mass \citep{iwamoto_1a}.  
This is thought to strongly constrain the minimum timescale over which SN~Ia can occur as white dwarfs are the culmination of low-mass stellar evolution (the remnants of stars with $M<8~{\rm{M}_\odot}$). 
Consequently, the minimum time taken for SN~Ia to explode is governed by a combination of the main sequence lifetime of the most massive WD progenitors ($\tau\sim100~\rm{Myr}$; \citealp{matteucci-ias}) and the dynamical evolution of the binary systems in a given stellar population.  
Thus, following an episode of star formation, it is expected that there is a characteristic delay before SN~Ia detonations begin, and a further timescale over which they explode in {\textit{significant}} numbers.  The distribution of SN~Ia delay times (SN~Ia DTD) for an SSP \citep[$R_{\rm Ia}$, e.g.][]{graur-dtd} is generally thought to scale as $\propto t^{-1}$ after a minimum delay \citep[see, e.g.,][]{Maoz-Graur-Alpha}, evidenced by the observed volumetric SN~Ia rate density.  SN~Ia are significant contributors to the Fe budget of the ISM \citep[see Fig. 2 of][for example]{wiersma-chemistry}, but synthesise no $\upalpha$-elements.

Analytical models of chemical evolution typically implement an exponential form of the SN~Ia DTD ($R_{\rm Ia} \propto e^{-t/\tau_{\rm Ia}}$ where $\tau_{\rm Ia}$ is the e-folding timescale) and a minimum delay time \citep[$t_{\rm d}$, e.g. 150 Myr per][]{nidever-lazy-giants,vice1,vice2,vice3}. 
As these models do not account for the hierarchical assembly of galaxies from their progenitors, by construction SN~II alone contribute to the chemical enrichment of the ISM over a period $t_{\rm d}$ following the onset of star formation in a nascent galaxy, leading to the formation of a flat sequence of elevated $[\upalpha/\rm{Fe}]$ at low $[\rm{Fe/H}]$, whose absolute value in [$\upalpha/\rm Fe$] is initially set by the [${\rm \upalpha/Fe}$] of pure SN~II yields. 
When SN~Ia enrichment begins, [$\upalpha$/Fe] progressively decreases, increasingly so as $t$ approaches the DTD's e-folding timescale. This results in [$\upalpha$/Fe] decreasing as [Fe/H] increases \citep[e.g.,][]{matteucci-chemistry}. The ensuing decrease in [$\upalpha$/Fe] leads to the appearance of a characteristic transition in the $\upalpha$-Fe plane commonly referred to as the ``$\upalpha$~knee''.  

For the purposes of this paper, we label the {sequence of approximately flat [$\upalpha$/Fe] as a function of [Fe/H] located at [Fe/H]<[Fe/H]$_{\rm{knee}}$ as the ``plateau'' and the sequence of declining [$\upalpha$/Fe] as a function of [Fe/H] located at [Fe/H]>[Fe/H]$_{\rm{knee}}$ the ``shin''}\footnote{The terms plateau and shin were originally coined by \cite{McWilliam_MW} and \cite{nidever-lazy-giants}, respectively.}. To date, a knee has been observed in a number of the Local Group Dwarfs (LGDs) \citep[e.g.][]{gonzales-11-afe, deboer-sag-knee, hendricks-knee, carlin-explosive-alpha, hasselquist-dwarfs} and the Galactic thick disk \citep[e.g.,][]{wallerstein_62, Bensby2014,hayden-2015,mackereth-accreted}.

There has been much interest in measuring the position of this knee in [$\upalpha$/Fe]-[Fe/H] space (respectively, [$\upalpha$/Fe]$_{\rm{knee}}$ and [Fe/H]$_{\rm{knee}}$).  Some studies \citep{hendricks-knee, nidever-lazy-giants} show that [Fe/H]$_{\rm{knee}}$ scales with both the stellar mass and dynamical mass of a galaxy, which is in broad agreement with observations of the $\upalpha$-abundances of massive galaxies \citep[e.g.][]{Worthey1992,Trager2000,Schiavon2007,Johansson12,Conroy_etg_14,segers-alpha}.  
Based on one-zone galaxy chemical evolution (GCE) models, some groups \citep[e.g.,][]{andrews} have interpreted the broad empirical evidence as resulting from a correlation between galaxy mass and the gas consumption timescale\footnote{The gas consumption timescale, $t_{\rm{g}}$, is the inverse of what is often termed the `star formation efficiency' (SFE) by the GCE community. We do not use this terminology in order to avoid possible confusion with the efficiency per free-fall time term in the Schmidt star formation law.}. 
If a galaxy's reservoir of star-forming gas is initially characterised by a short $t_{\rm{g}}$, more stars form before the minimum delay time of the SN~Ia DTD, and many more before the e-folding timescale $\tau_{\rm{Ia}}$. 
Thus, more stars form and experience chemical enrichment with the metallicity-dependent yield closer to that of SN~II {\it only}, leading to the formation of a longer plateau, and thus leading to a greater [Fe/H]$_{\rm{knee}}$.  Feedback efficiency is also likely to be important, as efficient feedback may expel much of the enriched gas in galaxy-wide outflows before it is incorporated into new stellar generations.  
Thus, the properties of the underlying dark matter distribution must be important as stellar feedback is less efficient in removing the star-forming gas in a galaxy with a deeper potential well and a high entropy \citep[e.g.][]{keller-14,davies-lowess_1}.

However, a growing number of observations \citep[e.g.,][]{vargas1-mw,vargas2-and, nidever-lazy-giants} suggest that the $\upalpha$~knee is not ubiquitous.  Measurements of [Fe/H]$_{\rm{knee}}$ in dwarf galaxies by \cite{hendricks-knee} and \cite{nidever-lazy-giants} show that the relationship between $M_\star$ and [Fe/H]$_{\rm{knee}}$ is subject to a large degree of scatter at fixed $M_\star$.  
In addition, there are dwarfs in the Local Group that display monotonically-decreasing trends of [$\upalpha$/Fe] as a function of [Fe/H], with no evidence for the presence of a knee \citep[e.g.,][]{vargas1-mw,vargas2-and}.  
Moreover, the premise that $\upalpha$~knee formation corresponds only to the onset of SN~Ia has been called into question by other authors.  On the basis that chemical enrichment by massive stars is highly correlated with ongoing star formation, \cite{tolstoy-09} suggested that a declining-$\upalpha$ shin could form due to a decrease in the SFR, which induces a prompt reduction of the SN~II rate but a delayed reduction in that of SN~Ia.


The emergence of cosmological simulations of galaxy formation that broadly reproduce a wide range of key properties of the observed present-day galaxy population \citep[for a recent review see][]{Crain_and_van_de_Voort_23} presents a timely opportunity to examine the predicted element abundances of galaxies modelled with fewer ad-hoc components than is the case for GCEs. Here we examine the $\upalpha$-Fe planes of a reasonably representative sample of present-day galaxies that form in a high-resolution cosmological hydrodynamical simulation of a periodic volume from the EAGLE project. 
The EAGLE simulations \citep{schaye-eagle, crain-calibration}, and simulations based on the EAGLE model, have been used to explore many related lines of enquiry, including the $\upalpha$-enhancement of massive galaxies \citep{segers-alpha}, the $\upalpha$-bimodality of the Milky Way disk \citep{mackereth-bimodality}, the radial gradient of oxygen abundances of galaxy discs \citep{tissera-oxy}, the $\upalpha$-enrichment of globular cluster populations \citep{hughes-gc}, the chemical composition of satellites accreted early in the Milky Way history \citep{horta-heracles}, and the influence of star formation histories on $\upalpha$-abundances \citep{gebek-alpha-eagle}.

Our chief aims are the following: {\it (i)} to establish the physical mechanism responsible for the formation of $\upalpha$~knees; {\it (ii)} to determine whether the $\upalpha$~knees are ubiquitous in well-resolved simulated galaxies; {\it (iii)} to determine what influences the position of the $\upalpha$~knee in this chemical space, given the fixed choices of IMF, SN~Ia DTD, stellar yields and lifetimes adopted by the simulations; {\it (iv)} to determine whether the simulations predict the existence of a [Fe/H]$_{\rm{knee}}$-$M_\star$ relation (hereafter, MKR), whose slope and scatter are consistent with observational constraints; and {\it (v)} to establish the physical processes responsible for the existence of the MKR and its associated scatter.

In $\S$\ref{sec2_data_method} we describe the simulation and the procedure we use to characterise the $\upalpha$-Fe planes of its galaxies.  In $\S$\ref{sec3_results} we characterise the broad properties of the galaxy population in terms of their $\upalpha$-abundance patterns, determine what forms the $\upalpha$~knee in the simulated galaxies and then examine the relationship between $\upalpha$~knee metallicity and stellar mass. In $\S$\ref{sec4_knee} we examine the element abundance evolution inferred by a one-zone GCE model for a fixed star formation history, similar to that of a simulated galaxy that exhibits a knee. We compare the outcomes using i) the constant gas consumption timescale assumed by default by the GCE, and ii) an evolving gas consumption timescale similar to that exhibited by the simulated galaxy, which is initially significantly longer than the default timescale. Using the latter, we find that the GCE closely reproduces the element abundance evolution of [O/Fe]$(t)$ and [Fe/H]$(t)$ exhibited by the simulated galaxy. We summarise our findings in $\S$\ref{sec5_summary}.

\section{Methods}\label{sec2_data_method}
 In $\S$\ref{sec2_1_EAGLE} we provide a brief overview of the EAGLE simulation used here. The simulations have been described in detail in many papers, so we present only a brief overview and restrict discussion of the subgrid models to those most relevant to this study. In $\S$\ref{sec2_2_dmo} we describe how haloes and galaxies are identified. In $\S$\ref{sec2_3_method} we present our method for quantitatively characterising the `morphology' of the $[\upalpha/\rm{Fe}]$-[Fe/H] distribution of simulated galaxies.

\subsection{The EAGLE simulations}\label{sec2_1_EAGLE}
The EAGLE simulations \citep{schaye-eagle, crain-calibration} are hydrodynamical simulations of the formation and evolution of galaxies in a $\Lambda$CDM cosmogony, adopting the parameters inferred by \cite{planck_results} ($\Omega_{0}=0.307$, $\Omega_{\rm b}=0.04825$, $\Omega_\Lambda=0.693$, $\sigma_{\rm s}=0.8288$, $n_{\rm s}=0.9611$, $h=0.6777$ and $Y=0.248$). The simulation data can be downloaded by the community from an online database, as described by \citet{mcalpine-data-release}. The simulations were evolved using a modified version of the $N$-body Tree-PM smoothed particle hydrodynamics (SPH) code GADGET-3 \citep[last described by][]{springel-gadget}.  The numerical modifications include the pressure-entropy SPH formulation of \cite{hopkins-sph}, the time-step limiter of \cite{durier-timestep}, and switches for artificial viscosity \citep{cullen-inviscid} and artificial conduction \citep{price}.  

The suite of subgrid models includes element-by-element radiative cooling and heating for 11 species \citep{wiersma-photocooling}; treatment of the ISM as a single-phase medium, subject to a polytropic pressure floor, that becomes eligible for stochastic conversion into star particles at a pressure-dependent rate that reproduces the observed \citet{KS-98} star formation law \citep{schaye-ks-law}, when the gas becomes denser than a metallicity-dependent density threshold \citep{schayethreshold}.  Star particles are treated as simple stellar populations (SSPs) with a \citet{chabrier_IMF} initial mass function (IMF) between 0.1 and 100 $\rm{M}_\odot$, whose stars evolve and lose mass following the model of \cite{wiersma-chemistry}.  This chemodynamics model uses the metallicity-dependent nucleosynthetic yields for massive stars, Type Ia SN, Type II SN and the asymptotic giant branch (AGB) phase \citep{portinari-yields,marigo-yields} and metallicity-dependent stellar lifetimes \citep{portinari-yields}.  The ‘lifetimes’ of SN~Ia are specified by an empirically-motivated exponential delay time distribution, such that their rate per unit initial stellar mass is
 \begin{equation}
     \label{eq1_r1a}
     \dot{N}_{\rm{SN~Ia}}(\tau) = \nu \frac{{\rm e}^{-t/\tau}}{\tau},
 \end{equation}
\noindent where $\nu=2\times10^{-3}\rm{M}^{-1}_{\odot}$ is the number of SN~Ia per unit initial stellar mass, and $\tau=2~\rm{Gyr}$ is the characteristic e-folding time.  These parameters were calibrated to ensure that the simulations broadly reproduce the observed evolution of the cosmic Type Ia SN rate density \citep{schaye-eagle}.  The EAGLE DTD implements a metallicity-dependent minimum delay time of $t_{\rm{d,~min}}\approx 40~\rm{Myr}$ (which is consistent with the shortest delay time of \citealp{matteucci-ias}; by contrast \citealp{schaye_flamingo} set the SN~Ia rate to zero below ages of exactly 40 Myr). This metallicity dependence follows from the metallicity dependence of the stellar mass-age relation (Schaller, private communication). It is worth noting that in the GCE modelling community, a longer minimum delay time of $t_{\rm{d,~min}}=150~{\rm{Myr}}$ is typically adopted.

Star particles inject feedback energy associated with the evolution and explosion of massive stars by stochastically and isotropically heating neighbouring gas particles by a temperature increment of $\Delta T_{\rm SF} = 10^{7.5}\,{\rm K}$ \citep{schaye-thermal-feedback}.  Black holes (BHs) of initial mass $10^5\hMsun$ are seeded in haloes with mass greater than $10^{10}\hMsun$, identified on-the-fly with a periodically-triggered friends-of-friends (FoF) algorithm.  BHs act as “sink” particles that grow through BH-BH mergers and Eddington-limited Bondi-Hoyle accretion, modulated by the circulation speed of gas close to the BH \citep{rosas-2015, bower17}.  Feedback energy associated with BH accretion is injected by stochastically and isotropically heating neighbouring gas particles by a temperature increment of $\Delta T_{\rm AGN} = 10^{9}\,{\rm K}$ \citep{booth2009,schaye-eagle}.

As our study principally concerns the detailed element abundance patterns of $\lesssim L^\star$ galaxies, superior resolution and particle sampling is desirable.  We therefore examine the Recal-L034N1034 simulation from the EAGLE suite \citep[first introduced by][]{bastian-emosaics}, which follows the evolution of a periodic cube of side length $L=34.4\,\rm{cMpc}$ populated with $1034^3$ collisionless dark matter particles of mass $m_{\rm DM}=1.21\times10^6~\rm{M}_{\odot}$, and an initially equal number of gas particles of initial mass $m_{\rm g}=2.26\times10^5~\rm{M}_{\odot}$. The simulation adopts a Plummer-equivalent gravitational softening length of $\epsilon_{\rm com}=1.33~\rm{ckpc}$, limited to a maximum proper length of $\epsilon_{\rm prop}=0.35~\rm{ckpc}$. The mass (spatial) resolution of the simulation is therefore superior to that of the flagship Ref-L100N1504 by a factor of 8 (2) and, as discussed by \citet{schaye-eagle}, at this higher resolution it was necessary to recalibrate the efficiency of the subgrid stellar feedback model's parameters in order to achieve a satisfactory reproduction of the present-day galaxy stellar mass function (GSMF). The Recal-L034N1034 simulation used here has a volume of $\simeq 2.6$ greater than the Recal-L025N0752 simulation introduced by \citet{schaye-eagle}. It also incorporates the E-MOSAICS treatment of star cluster formation and disruption \citep{pfeffer-emosaics,kruijssen-emosaics}, though we do not examine the properties of star clusters here.

We adopt the standard definition of the abundance ratio of elements $x$ and $y$ relative to their Solar abundance ratio:
\begin{equation}\label{eq_abundances}
    \left[\frac{x}{y}\right] = \log_{10}\left(\frac{X^x}{X^y}\right) - \log_{10}\left(\frac{X^x_\odot}{X^y_\odot}\right),
\end{equation}
where $X^x = \sum_i m_i^x/\sum_i m_i$ is the mass fraction of stellar particle $i$ in element $x$, and $m_i$ and $m_i^x$ are the total mass of the stellar particle, and its mass in element $x$, respectively. We adopt the Solar abundances of \cite{asplund-eagle-solar}, in which $X^{\rm O}_\odot/X^{\rm Fe}_\odot = 4.44$. We follow \cite{segers-alpha} and use the abundance ratio [O/Fe] as a proxy for [$\upalpha$/Fe], because O dominates the mass fraction of the $\upalpha$-elements.

\begin{figure*}
	\includegraphics[scale=0.71]{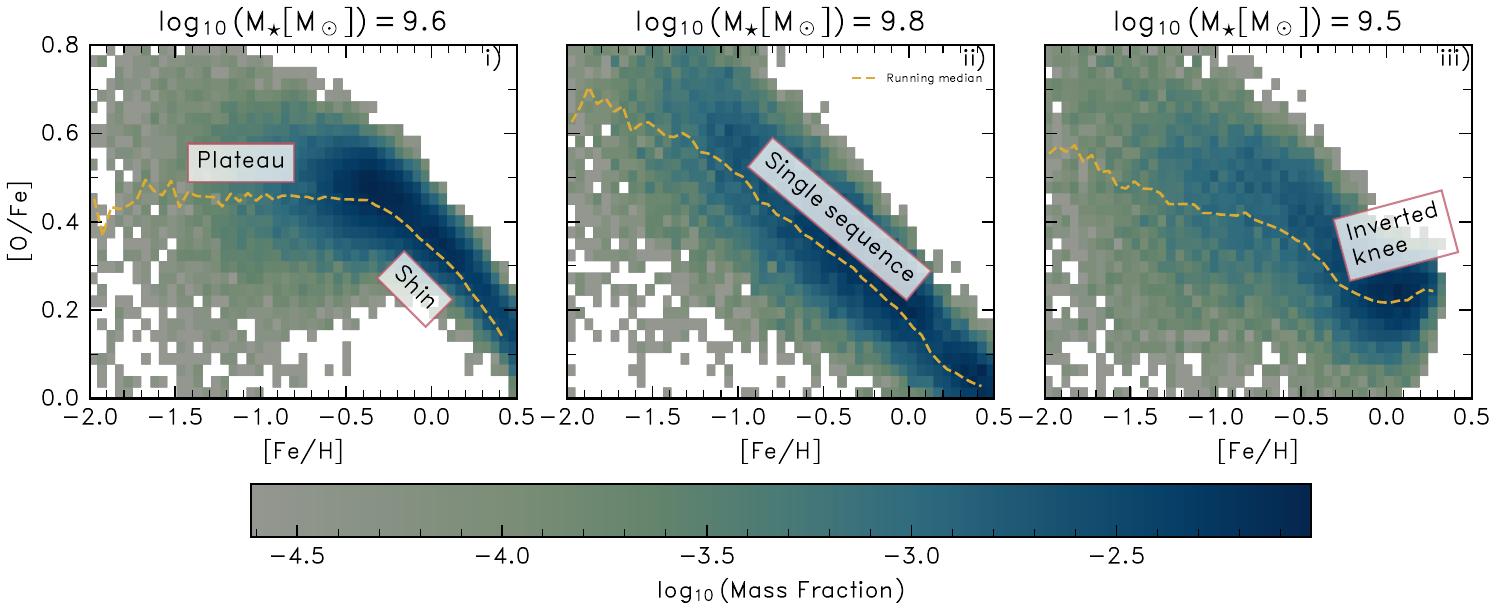}
    \caption{The $\upalpha$-Fe planes of three present-day galaxies from the simulation, selected as exemplars of the main categories of $\upalpha$-Fe distribution we find, represented as initial mass-weighted 2-dimensional histograms of stellar abundances. Overlaid dashed yellow curves denote the median of [$\upalpha$/Fe] as a function of [Fe/H]. From left to right, we show {\textit{i)}} a classical knee, whose $\upalpha$-Fe plane exhibits the canonical plateau-knee-shin morphology, {\textit{ii)}} a single-slope whose $\upalpha$-Fe plane is characterised by a single declining sequence of [$\upalpha$/Fe] as a function of [Fe/H] and {\textit{iii)}} an inverted knee, characterised by an initial declining sequence of [$\upalpha$/Fe] as a function of [Fe/H] before a turnover to a positive gradient at high [Fe/H]. A mass fraction of $\log_{10}(f_{\rm bin})=-3$ corresponds to a stellar mass per pixel of $M_{\star,{\rm bin}}=(3.6, 6.5, 3.1) \times10^{6}~{\rm{M}}_\odot$ for the left, middle and right panels, respectively.}\label{fig_threepanel_afe}
\end{figure*}

\subsection{Halo and galaxy finding}\label{sec2_2_dmo}
Galaxies and the haloes that host them are identified by applying the FoF algorithm to the dark matter particle distribution with a linking length 0.2 times the mean interparticle separation.  Gas, star and BH particles are then associated to the FoF group of the nearest dark matter particle.  Substructure bound to these haloes is then identified using the $\textsc{SUBFIND}$ algorithm (\citealp{springel-subfind}, \citealp{dolag-subfind}). In each FoF group, the subhalo that contains the most bound particle is designated the central galaxy/subhalo; the other subhalos of the FoF group are satellites. When referring to the aggregated properties of the gas, star, and dark matter particles comprising a galaxy (for example, quantities such as a galaxy's stellar mass, star formation rate or metallicity), we follow \citet{schaye-eagle} and compute the properties from particles within a spherical aperture of comoving radius $r=30~{\rm ckpc}$ centered on the most bound particle. We examine only galaxies with a present day stellar mass of $M_\star > 10^{8.5}\,{\rm{M}_\odot}$, ensuring that galaxies are sampled by $\gtrsim 2000$ star particles. 

\subsection{Characterising simulated \texorpdfstring{$\upalpha$}{alpha}-Fe distributions}\label{sec2_3_method}

Inspection of the $\upalpha$-Fe planes of the simulated galaxies reveals that there is significant diversity among the galaxy population at $z=0$. Fig.~\ref{fig_threepanel_afe} provides illustrative examples of the three main categories of $\upalpha$-Fe plane morphology that we find, from left to right i) galaxies that exhibit the canonical plateau-knee-shin shape, ii) a single sequence of declining [$\upalpha$/Fe] and [Fe/H] and iii) galaxies with an increasing $\upalpha$-element abundances at high [Fe/H]. Overlaid dashed yellow curves denote the median [O/Fe] as a function of [Fe/H], computed in bins of $\Delta$[Fe/H]=0.05. We stress that this median does not necessarily reflect in detail the temporal evolution of the ISM element abundances since, for example, the accretion of a gas-rich progenitor can briefly lower [Fe/H].

The diversity shown in the $z=0$ galaxy population in Fig.~\ref{fig_threepanel_afe} illustrates the need for a flexible procedure to quantitatively characterise these $\upalpha$-Fe morphologies.  Therefore, we model them as a two-piece piecewise linear function, parameterised as a function $f([\rm{Fe/H}]_{knee},~[\upalpha/\rm{Fe}]_{knee},~\theta_{1},~\theta_{2})$ where $\theta_1$ and $\theta_2$ are, respectively, the slopes of the plateau and the shin component. \cite{hendricks-knee} fitted a piecewise linear function to the abundances of the Sag dSph assuming that their trajectory was consistent with the canonical $\upalpha$-Fe track shown schematically in the review of \cite{McWilliam_MW} which comprises a high-[$\upalpha$/Fe] plateau, connected to a low-[$\upalpha$/Fe] plateau at high [Fe/H] by a declining $\upalpha$-shin.  However, this parameterisation was made on the assumption that {\textit{all}} galaxies should show an $\upalpha$~knee that conforms to the canonical morphology.  The evidence from the Local Group and our initial examination of the simulated galaxies motivates the relaxation of these constraints.  Therefore, we place no restrictions on the signs of the gradients of the two components, nor on the value of the breakpoint in ([O/Fe],[Fe/H]) space that separates them.  Thus, we can identify cases where the $\upalpha$-Fe plane shows a declining or increasing trend of [$\upalpha$/Fe] as a function of [Fe/H].  Both morphologies have been observed in a number of stellar populations in the Local Group: \cite{hayden-2015}, \cite{nidever-lazy-giants}, \cite{hasselquist-dwarfs},  \cite{horta-substructure}, and \cite{Fernandes2023} provide examples of positive gradients, and \citet{vargas1-mw, vargas2-and} provide examples of mono-gradient $\upalpha$-Fe planes.

To construct a statistical model of the $\upalpha$-Fe plane of individual galaxies, we use the Python package \textsc{pymc3} to fit a piecewise linear model to the data using Bayesian inference.  We follow the procedure outlined by \cite{hogg-pw} assuming that there are no measurement uncertainties on the abundances of the simulated galaxies. The best-fitting model is found by maximising the likelihood for the parameters of the piecewise model $O=[\rm{[Fe/H]}_{knee},\rm{[O/Fe]}_{knee}, \theta_1, \theta_2]$ which we assume takes the form:

\begin{equation}
    \label{eq2_log_likelihood}
    \ln\mathcal{L}(O|\rm{[Fe/H], [O/Fe]}|) = K - \sum\limits_{i=1}^{N}(\frac{\Delta^2_i}{2\Sigma^2_i}+\ln|\Sigma^2_i|)
\end{equation}

\noindent where $\Delta^2_i$ defines the distance between the $i^{\textit{th}}$ data point and the model and $\Sigma^2_i$ is the variance orthogonal to the model.  K is a normalisation constant.  We assume that the priors for the model parameters are Gaussian and that the constant scatter term for [O/Fe] at fixed [Fe/H], $\Sigma$, is described by a half-Cauchy prior.  After minimising the log-likelihood we use this optimal solution to initiate a Markov Chain Monte Carlo (MCMC) sampling of the posterior probability distribution function (PDF) of the parameters $O$ using the No-U-Turn MCMC Sampler \citep[NUTS; ][]{NUTS-sampler} as implemented in $\textsc{pymc3}$ \citep{pymc3}.  This produces an estimate of the posterior PDFs of each parameter from which we take the median values of $O$.  We take the interquartile range of each estimated posterior PDF to be the uncertainty of a given parameter.

\begin{figure}
    \includegraphics[scale=1]{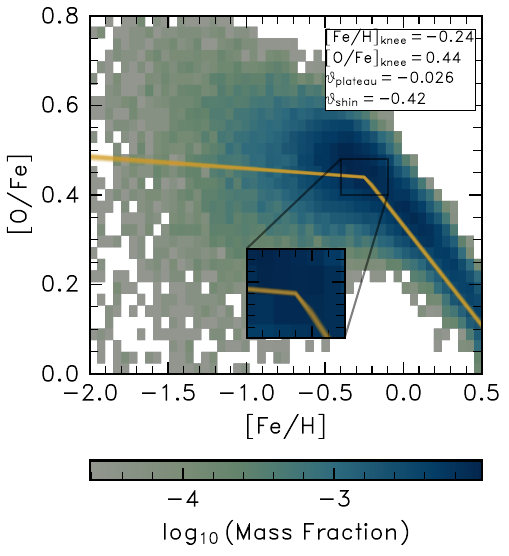}
        \caption{The initial mass-weighted histogram of the galaxy exhibiting a classical knee $\upalpha$-Fe distribution shown in panel i) of Fig.~\ref{fig_threepanel_afe}, shown here with the resultant piecewise linear fit determined by linear regression to the data. We superimpose 50 realisations of our piecewise model as sampled randomly from the posterior PDFs of each parameter over the range in [Fe/H], with an inset axis zooming in on $[\rm{Fe/H}]_{knee}$.}\label{fig_ck_fit}
\end{figure}

Fig.~\ref{fig_ck_fit} shows the 2d initial mass-weighted histogram of the distribution of [O/Fe] vs.  [Fe/H] for the classical $\upalpha$~knee shown in the left panel of Fig.~\ref{fig_threepanel_afe}, where the piecewise model is overlaid by solid a line. The interquartile range, which is very small, is shown by many transparent yellow curves which show 50 models with parameters drawn randomly from the posterior distributions inferred by the method.  For this galaxy we infer a value $[\rm{Fe/H}]_{knee}=-0.24\pm0.02$. As we show the running median in Fig.~\ref{fig_threepanel_afe}, it is less sensitive to stellar particles with high [O/Fe] at fixed [Fe/H] and thus appears flatter.

The simulated galaxies can be categorised into the classes illustrated by Fig.~\ref{fig_threepanel_afe}, based on the outcome of our fit to their $\upalpha$-Fe planes: {\it classical knee} distributions are those for which the fitted piecewise function clearly resembles an $\upalpha$-rich sequence or plateau, and a declining $\upalpha$ shin sequence, {\it single slope} distributions are dominated by a monotonically decreasing sequence of [O/Fe] vs. [Fe/H] with no obvious knee, and {\it inverted knee} distributions exhibit $\upalpha$-plateaus or even classical knees but transition at relatively high [Fe/H] to a sequence of increasing [$\upalpha$/Fe] as a function of [Fe/H].

For the stellar particles of galaxies with a classical knee, we define the plateau and shin stellar populations as those with [Fe/H] less than, or greater than, [Fe/H]$_{\rm{knee}}$, respectively.  This allows us to compute the aggregated properties of the stellar populations which comprise the plateau and shin.  In order to select a representative sample of classical knees, we impose the following selection criteria: 
    
\begin{enumerate}
    \item $\theta_{\rm{shin}}+\sigma_{\theta,~\rm{shin}}<\theta_{\rm{plateau}}$
    \item $\theta_{\rm{plateau}}-\sigma_{\theta,~\rm{plateau}}>\theta_{\rm{shin}}$
    \item  $M_{\star,{\rm plateau}}/M_{\star,{\rm ~total}}>0.25$
    \item$M_{\star,{\rm shin}}/M_{\star,{\rm ~total}}>0.25$,
\end{enumerate}

where $\theta$ refers to the slope of a component, $\sigma$ refers to the standard error (measured from the posterior distributions of each parameter) and $\rm{M}_{\star,\rm{plateau/shin/total}}$ refer to the plateau, shin and total stellar masses.  Our criteria for the gradients of each component of the piecewise model and on the masses of the plateau are to ensure that a change in slope and the masses of each component are statistically significant.

Galaxies in the single slope category are selected so that the slopes of the plateau and shin are the same within the measured errors such that:

\begin{enumerate}
    \item $\theta_{\rm{plateau/shin}}\leq0$
    \item $\theta_{\rm{shin}}+\sigma_{\theta,~\rm{shin}}>\theta_{\rm{plateau}}$
    \item $\theta_{\rm{plateau}}-\sigma_{\theta,~\rm{plateau}}<\theta_{\rm{shin}}$,
\end{enumerate}
are satisfied. 

\section{Results}\label{sec3_results}

In order to gain insight into the physics responsible for $\upalpha$~knee formation, we contrast the properties of galaxies that exhibit classical knees with those having a single slope $\upalpha$-Fe morphology.  In $\S$\ref{sec3_1_census}{Sec. 3.1} we characterise quantitatively the galaxies' $\upalpha$-Fe distributions. In $\S$\ref{sec4_knee} we examine their star formation and chemical enrichment histories, in order to identify mechanisms responsible for their markedly different $\upalpha$-Fe distributions. Finally, in $\S$\ref{sec_5_mkr} we validate the simulation's outcomes via a comparison to observations of stellar populations in the Local Group, focusing on the slope of the relationship between [Fe/H]$_{\rm{knee}}$ and $M_{\star}$ (dubbed `MKR' -- $\textbf{\textit{M}}_\star$-[Fe/H]$_{\rm \textbf{k}nee}$ \textbf{r}elationship) and its scatter. 

\subsection{A census of classical knees and single slopes in the simulated galaxy population}\label{sec3_1_census}

\begin{figure}
	\includegraphics{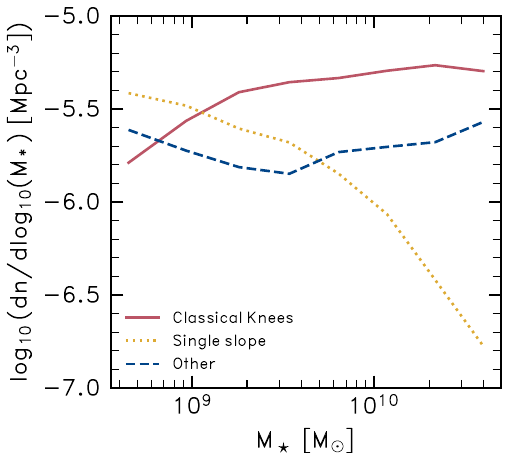}
    \caption{The fractions of the present-day galaxy population in the simulation, as a function of galaxy stellar mass, whose $\upalpha$-Fe distributions are categorised as `single slope', `classical knee' or `other'. At low mass, most of the galaxy population does not exhibit classical knees, but as $M_\star$ increases the fraction of single slopes decreases, with that of classical knees commensurately increasing. The fraction of bimodal, inverted knees and unclassified galaxies is largely insensitive to stellar mass.}\label{fig_frac_gals}
\end{figure}

Application of the criteria described in \S\ref{sec2_3_method} yields 540 simulated galaxies with classical knees, 310 with single $\upalpha$-Fe slopes and 241 whose $\upalpha$-Fe planes cannot be simply described by a two-piece piecewise linear function. 
The latter category includes both inverted knees and galaxies that   exhibit $\upalpha$-bimodality (a phenomenon that in this simulation is only seen at  $M_\star \simgreater 10^{10}~{\rm{M}_{\odot}}$; \citealp{mackereth-bimodality}), which together account for $\approx30$ percent of galaxies, irrespective of galaxy mass. 
Although the classical knees represent just under half the overall sample, there is a significant mass dependence: Fig.~\ref{fig_frac_gals} reveals that for $M_\star\lesssim10^9~\rm{M}_\odot$ the dominant category of galaxy is, marginally, the single-slopes, with the fraction of galaxies exhibiting classical knees increasing monotonically with $M_\star$. 

\begin{figure}
	\includegraphics{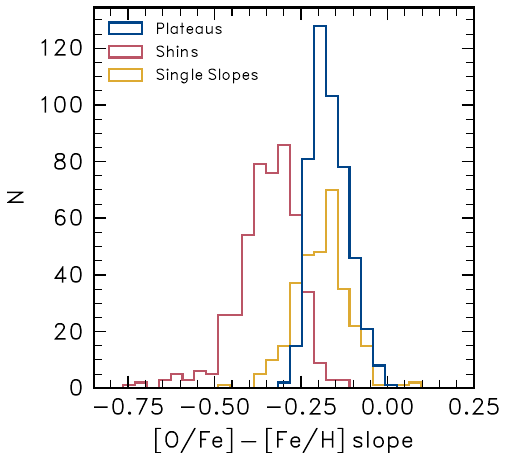}
    \caption{Histogram showing the characteristic slopes of {\textit{i)}} $\upalpha$ plateaus, {\textit{ii)}} $\upalpha$-shins and {\textit{iii)}} single-sloped $\upalpha$-Fe planes.  The distributions of these parameters show that the characteristic slope of the plateau is not close to zero - such slopes are rare in the simulation. Furthermore the slopes of $\upalpha$-plateaus and single slopes seem to be consistent, indicative of them having similar enrichment histories. However, the shins of classical knees are sharply declining by comparison.}\label{fig_gal_slopes}
\end{figure}

Fig.~\ref{fig_gal_slopes} shows the distribution of the gradients of the sequences recovered from our characterisation of the $\upalpha$-Fe planes of the simulated galaxies: that of the overall sequence in single-slope galaxies, and those of the plateau and shin components of classical knee galaxies. 
For the latter, we find typical ratios in the gradients of the two sequences of a factor $\simeq 2$. The plot also reveals that the typical gradient of the overall sequence in single-slope galaxies is similar to that of the plateau, and not that of the shin of classical knee galaxies. 

These results are in conspicuous contrast to predictions of chemical evolution models in which  early evolution results in a horizontal plateau at the [$\upalpha$/Fe] value corresponding to the IMF-averaged yield of Fe and $\upalpha$-elements resulting from {\textit{pure}} SN~II enrichment \citep[e.g.][]{thomas-imf-yields}.  
In that idealised case we would expect $\theta_{\rm{plateau}}=0$.  The clear difference between the slopes of plateaus and shins is a signature of differing modes of chemical enrichment.  Thus, we can immediately infer that in spite of these galaxies having extended histories of star formation, the formation of single-sloped galaxies must in some way resemble that of high-$\upalpha$ plateaus.  
This would then imply that the occurrence of $\upalpha$-knees and the extent of the plateau in [Fe/H] are not related solely to the form of the DTD. 

We are thus left with two conclusions to draw from Fig.~\ref{fig_frac_gals} \& Fig.~\ref{fig_gal_slopes}.  Firstly, Fig.~\ref{fig_frac_gals} indicates that the single-sloped galaxies are reasonably common in the simulated galaxy population, particularly at low mass.  
Secondly, recalling that the EAGLE model adopts a SN~Ia DTD with a very short minimum delay time, we nevertheless find that more than half of the simulated galaxies exhibit Fe-rich $\upalpha$-knees. 
Such $\upalpha$~knees therefore cannot be solely caused by the onset of SN~Ia. Bearing in mind that $\upalpha$-plateaus and single slopes may have similar origins, we turn to the formation histories of classical knees and single slopes in the next sub-section, with the aim of identifying the physical driver(s) of their  $\upalpha$-Fe plane differences.

\subsection{The formation of \texorpdfstring{$\upalpha$}{alpha}-Fe knees}\label{sec4_knee}

In this sub-section, we compare the star formation and element abundance histories of classical knees and single-sloped galaxies in order to establish why the former show $\upalpha$~knees while the latter do not, and in particular to scrutinise the orthodox interpretation that $\upalpha$-knees form as a consequence of the {\it onset} of SN~Ia, whether this onset is defined by the minimum delay time or the e-folding timescale. 

\subsubsection{The star formation histories of galaxies with classical knees and single slopes}\label{sec41}

\begin{figure*}
	\includegraphics{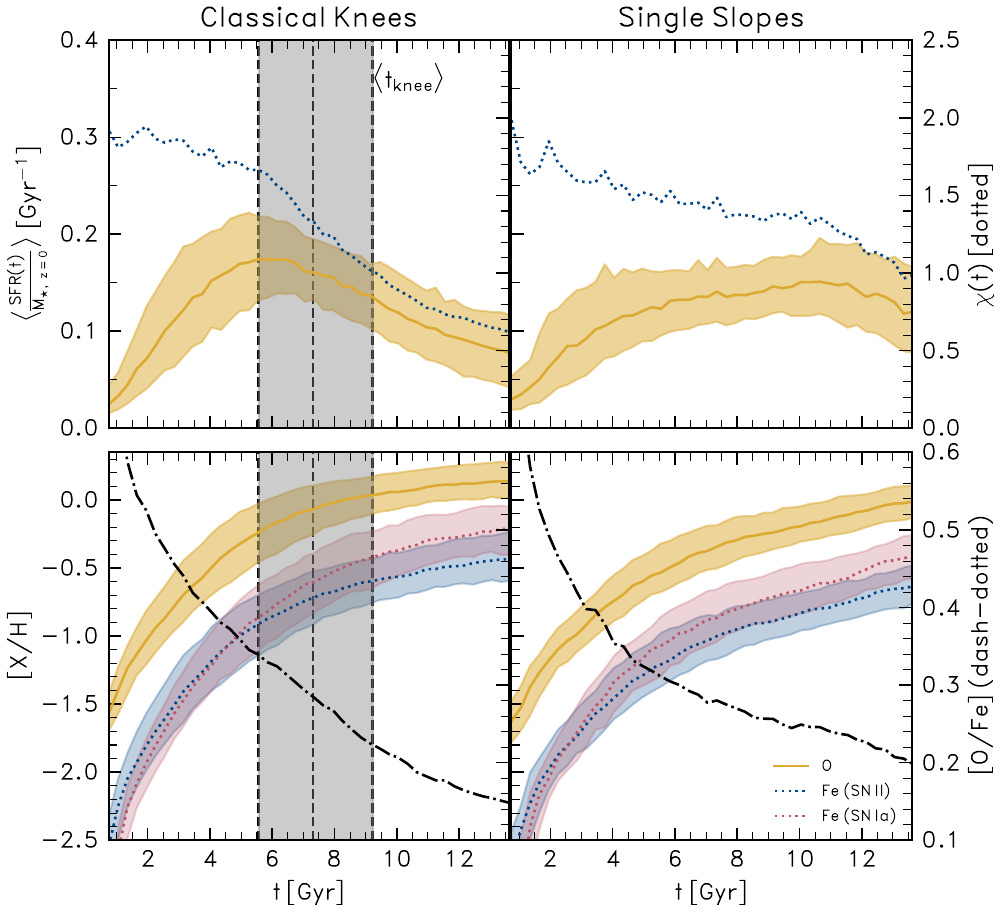}
    \caption{The star formation (top) and chemical enrichment (bottom) histories of galaxies exhibiting classical knees (left) and single slopes (right). The median star formation rate as a function of cosmic time is shown normalised by the present-day galaxy stellar mass, to compress the dynamic range. In the left panels the average time at which the $\upalpha$~knee forms is denoted by a black dashed line, with the shaded area representing the interquartile range.  The bottom panels show the evolution of [O/H] and [Fe/H], the latter decomposed into the contribution from SN~II and SN~Ia which are explicitly tracked in the simulations. The evolution of [O/Fe] is shown, relative to the right-hand axis, with a dashed-dotted line. Clear differences in star formation and chemical enrichment histories of the two categories of galaxy are evident. The SFHs of classical knees are characterised by a steep initial SFR increase followed by a decline around $t\simeq6~\rm{Gyr}$. At the time of $\upalpha$~knee formation, $\rm{[Fe/H]}_{SN~II}$ begins to plateau, whereas $\rm{[Fe/H]}_{SN~Ia}$ is increasing. The dotted lines in the top panels denote the instantaneous SFR divided by the past mean SFR (right y-axis). The time at which this track drops below unity coincides broadly with the formation of the $\upalpha$~knee. The formation of the $\upalpha$~knee is \textit{not} associated with merely the onset of SN~Ia but rather the contribution of SN~II diminishing as the $\rm{SFR}(t)$ declines.} \label{fig_4panel_singles_knees}
\end{figure*}

Fig.\,\ref{fig_4panel_singles_knees} shows the history of star formation and element abundance evolution of classical knees and single slopes. All evolutionary tracks are computed in 0.3 Gyr bins between $z=20$ to $z=0$ ($t\simeq0$ to $t\simeq13.8$), where instantaneous abundance ratios reflect the median abundances of newly-formed stellar particles in that interval.
The solid yellow curves in the top panels correspond to the median star formation histories (SFHs) of the classical knee and single slope samples (left and right panels, respectively). 
As our sample comprises galaxies with present-day stellar masses spanning $\simeq2.5$\,dex, we normalise the star formation rates by the present day stellar mass. This yields a quantity with unit $\rm{Gyr}^{-1}$, but we stress that this is not the same as the specific star formation rate for $z>0$. The shaded regions encompass the interquartile range of the SFHs. The SFHs of galaxies with classical knees are characterised by a steady increase to a peak of short duration: averaged over all classical knees in our sample, the peak occurs at $t\simeq7.5$ Gyr but, owing to cosmic downsizing \citep[e.g.][]{cowie-downsizing,qu_gama_17}, the peak tends to be found earlier (later) for more (less) massive galaxies.
Following the peak, the SFH experiences a sustained decline to the present day. Conversely, single-sloped galaxies exhibit SFHs that rise more steadily and remain reasonably constant over a Hubble time. 

The dotted lines in the top panels show the ratio between the ongoing SFR (i.e., at time $t$, averaged over the preceding 300~Myr), and the past average SFR up until that time. We denote this quantity $\chi$. 
The relative rates of SN~II and SN~Ia reflect the ratio of current-to-past histories of star formation \citep{gilmore_91_sne}. Thus a declining value of $\chi(t)$ indicates that evolved stellar populations are becoming more relevant to the chemical evolution, and a value of $\chi < 1$ implies that evolved stellar populations are beginning to dominate the chemical enrichment of Fe. In Fig.~\ref{fig_4panel_singles_knees} we see that when the peak occurs in the classical knee SFH plot, $\chi(t)$ drops below unity while the approximately constant SFR of the single slopes results in a track of $\chi(t)>1$ at all times.

The grey shaded region enclosed by dashed vertical lines in the left panels denotes the typical time at which the $\upalpha$~knee forms for these galaxies, $t_{\rm{knee}}$, measured directly from the time at which [Fe/H]$(t)$ crosses $[\rm Fe/H]_{knee}$. We see that it generally occurs $\simeq2~\rm{Gyr}$ after the peak SFR is reached. 
Therefore, the formation of the $\upalpha$~knee generally precedes the time at which $\chi$ first drops below unity, which is highly suggestive that evolved stellar populations begin to dominate the Fe-enrichment of the ISM during the evolution of the ``shin'' phase.  Recalling that the EAGLE model implements $R_{\rm Ia}$ with a short (40 Myr) minmum delay time and e-folding timescale of 2~Gyr, it is clear that the formation of the $\upalpha$~knee cannot be attributed to the `onset' of SN~Ia beginning to contribute to the chemical enrichment, whether this is defined as the minimum delay time or the e-folding timescale of the DTD.

\subsubsection{The effects of differing SFHs on galaxy element abundance evolution}\label{sec42}

The curves in the bottom panels of Fig.\,\ref{fig_4panel_singles_knees} show the evolution of the characteristic element abundances of new stellar populations, as a function of cosmic time, in the two categories of galaxy. 
The panels show the oxygen abundance [O/H] (solid yellow curves), and the iron abundances contributed separately by the two types of SN tracked by the simulation ([Fe/H]$_{\rm SN~II}$, dotted blue curves) and SN~Ia ([Fe/H]$_{\rm SN~Ia}$ dotted red curves). This is afforded by the fact that the simulations explicitly track the fraction of Fe released into the ISM by SN~Ia.
 For reference, the $\upalpha$-richness of new stellar populations is shown, relative to the right-hand axis, with a dot-dashed line. Unsurprisingly, SN~II dominate the initial Fe enrichment, but in both categories of galaxy SN~Ia eventually dominate, despite yielding slightly less than half of the iron of evolved stellar populations in EAGLE \citep{wiersma-chemistry}. 
We attribute this to the preferential ejection of SN~II nucleosynthetic products from the ISM in feedback-driven outflows. 

The differing star formation histories of the classical knee and single slope galaxy populations imprint significant differences on the evolving iron abundances of their new stellar populations. In single slope galaxies the ratio [Fe/H]$_{\rm SN~Ia}$/[Fe/H]$_{\rm SN~II}$ remains fairly constant, as denoted by the parallel red and blue dotted tracks in the bottom right panel of Fig.\,\ref{fig_4panel_singles_knees}.
In classical knee galaxies, following the peak of the SFR, the contribution of SN~II to the increasing iron abundance of newly-formed stars declines, as denoted by the flattening of the dotted blue curve. 
In contrast, SN~Ia continue to contribute significantly, as their ongoing rate does not react promptly to the decline of the SFR. Over the course of several Gyr, the [Fe/H]$_{\rm SN~Ia}$ and [Fe/H]$_{\rm SN~II}$ curves diverge, leading to the steady evolution of the gradient of the median track through [$\upalpha$/Fe]-[Fe/H] space, from that of the plateau to that of the shin. This divergence is likely exacerbated by the evolving efficiency with which Fe from the two types of SN is retained in the ISM and subsequently locked up in stars: prior to the decline of the SFR in classical knee galaxies, Fe released into the ISM by SNe~Ia can be entrained in outflows driven by SNe~II from ongoing star formation, just as is the case for single slope galaxies. After the SFR declines, the outflows also decline and Fe from SNe~Ia is more readily retained in the ISM.


\begin{figure}
	\includegraphics{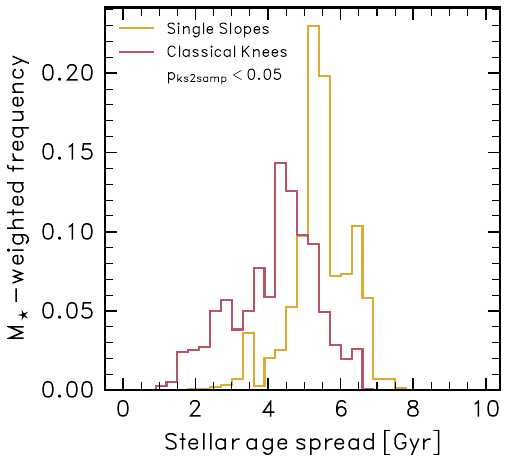}
    \caption{The characteristic age spread (i.e. interquartile range) of the stellar populations comprising simulated galaxies that exhibit $\upalpha-$Fe planes with a single slope (green) and that exhibit classical knees (red). As one might expect on the basis of Fig.~\ref{fig_4panel_singles_knees}, the single slope distribution shows a peak at a characteristically greater spread in stellar ages than that of the classical knees. An observational signature of the $\upalpha$~knee, then, might be a stellar age spread which could be measured via stellar population synthesis using the integrated light of unresolved stars.}\label{fig_age_spread}
\end{figure}

Based on Fig.~\ref{fig_4panel_singles_knees}, it is clear that classical knees and single slopes are distinguished by their differing star formation histories. The peaked star formation histories of classical knees should therefore yield a narrower spread of stellar ages in comparison to those resulting from the relatively constant SFR of the single slope galaxies. 
We measure the age spread of the stars in the two types of galaxies by equating it to the interquartile range of a galaxy's stellar ages. The resulting histograms (normalised by the total number of galaxies in each category) are plotted in Fig.~\ref{fig_age_spread}. 
The plot shows that as a  consequence of their extended SFHs, single slope galaxies indeed tend to have larger age spreads: on average 5.4 Gyr as opposed to the 4.2 Gyr of classical knees. 
We perform a two-sample Kolmogorov-Smirnov test and, obtaining a value of $p\ll0.05$, reject the null hypothesis that the age spread distributions are drawn from the same parent distribution.  Consequently, we conclude that, in principle, the presence of an $\upalpha$~knee could be inferred in a galaxy in so far as such an age spread can be constrained observationally for the unresolved stellar populations of more distant systems.

\subsubsection{Implications for the emergence of \texorpdfstring{$\upalpha$}{alpha}~knees}
\label{sec:interpretation}

It is widely assumed that the $\upalpha$~knee is caused by the $\textit{onset}$ of SN~Ia contributing significant Fe mass to the ISM {(e.g., \citealp{tinsley-chemistry}; \citealp{gilmore_91_sne})}.  Indeed, an early $\upalpha$~knee,  following a horizontal plateau at very low [Fe/H] {\textit{may}} appear following the very first SN~Ia explosions, though this may not be the case depending on how strong the metallicity dependence of SN~II nucleosynthetic yields is.  However, the more significant $\upalpha$~knees identified in our simulated galaxies are located at sufficiently high metallicity that their formation cannot be associated with the emergence of such early SN~Ia events.  

Our explanation for the emergence of $\upalpha$~knees is well aligned with suggestions made in previous work. \cite{tolstoy-09} has qualitatively argued that the $\upalpha$~knees seen in the chemical abundances of the LGDs could originate from a decline in their SFRs, and further suggested that due to their low masses dwarf galaxies are particularly susceptible to sudden events reducing their SFRs.  The rate of enrichment by SN~II reacts more promptly to this decline than the rate of enrichment by SN~Ia, due to a combination of the short lifetime of massive stars and the time dependence of the SN~Ia DTD, which evolves on gigayear timescales.
The SN~II progenitors with the longest lifetimes are shorter-lived than the typically adopted minimum delay time for the SN~Ia DTD.  

Our finding has also been anticipated in the work by \cite{Maoz-Graur-Alpha}. Those authors convolved the cosmic SFH of \cite{madau_17_dtd} with a power-law parameterisation of the SN~Ia~DTD and re-estimated the SN~Ia rate density under the assumption of a \cite{kroupa} IMF. 
They used these forms of the SFH and DTD to compute the mean $\upalpha$ and Fe accumulation history that results from the cosmic SFH. The evolutionary tracks of [$\upalpha$/Fe] and [Fe/H] produced by this model yield an $\upalpha$-plateau that gently declines until $z\lesssim2$ whence an $\upalpha$~knee forms due to a combination of the steep decline of the cosmic SFR (thus reducing the SN~II rate) and the shallower decline of the SN~Ia rate due to the power-law tail of the DTD. By contrasting Fig. 3 of \cite{schaye-eagle} to Fig. 1 of \cite{Maoz-Graur-Alpha}, which show model SN~Ia rate densities compared to the same set of DTD measurements, we can see that this behaviour is qualitatively mirrored by the exponential DTD of the EAGLE simulations.

\bigskip

To summarise, in the simulation examined here the formation of an $\upalpha$~knee is neither ubiquitous nor associated merely with the onset of SN~Ia enriching the star-forming gas of a galaxy.  Instead, a distinct change of gradient in the $\upalpha$-Fe plane is associated with a decline in the SFR, which brings about a transition from chemical enrichment being dominated by young stellar populations to old stellar populations.  As a natural consequence, we find that single slopes host on average younger stellar populations with a wider spread in age than their classical knee counterparts.

\subsection{The \texorpdfstring{$[\rm{Fe/H}]_{knee}$-$M_\star$ relation}{Knee metallicity-Mstar relation}}\label{sec_5_mkr}

Having determined how the $\upalpha$~knee forms in the simulated galaxies, we now turn to the relationship between the metallicity of the $\upalpha$~knee ([Fe/H]$_{\rm{knee}}$) and total stellar mass ($M_{\star}$) of this subset of galaxies.  We term this relationship the `MKR' (Mass-Knee metallicity Relation), and compare it on a quantitative basis to that measured using spectroscopic abundances and stellar mass estimates for the stellar populations comprising the galaxies of the Local Group.  We then examine the origin of the $\Delta$[Fe/H]$_{\rm{knee}} \simeq 0.8$ scatter at fixed $M_\star$ that emerges in the simulation, and discuss it in the context of the LGD observations.

\begin{figure}
	\includegraphics[width=\columnwidth]{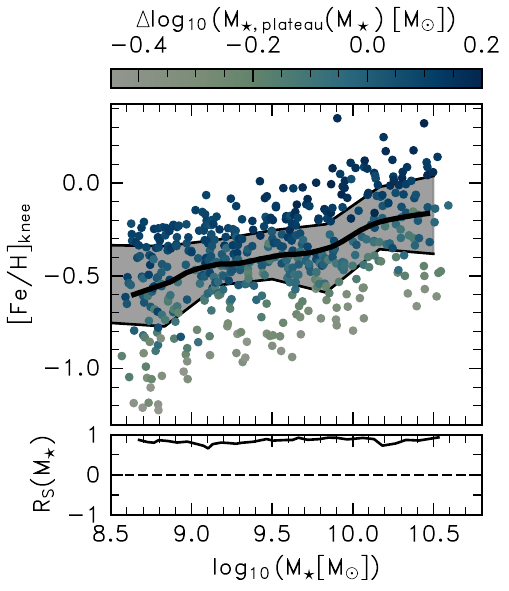}
    \caption{Scatter plot of the relationship between the $\upalpha$~knee metallicity, $\rm{}[Fe/H]_{knee}$, and the stellar mass of classical knee galaxies (the `mass - knee metallicity relation'). Each circle symbol represents an individual simulated galaxy.  The black curve and grey shared region denote, respectively, the binned median and interquartile range of $\rm{}[Fe/H]_{knee}$ as a function of $M_\star$. Symbols are coloured by the residuals of the relationship between $M_\star$ and $M_{\star,~\rm{plateau}}$ in moving windows of `fixed' $M_\star$. This shows that at fixed M$_\star$ there is a correlation between [Fe/H]$_{\rm{knee}}$ and $M_{\star,~\rm{plateau}}$ which means that the scatter in the relationship is driven by variation in how long the star formation rate is rising, forming the plateau. The inset-axis shows the moving Spearman rank correlation coefficient ($R_S$) of the variables in moving windows of $M_\star$.}\label{fig_mkr_rs}
\end{figure}

In the top panel of Fig.~\ref{fig_mkr_rs} we show the MKR of the present-day simulated galaxies.  We find a monotonically increasing relation between $[\rm{Fe/H}]_{\rm{knee}}$ and $M_\star$.  
The Spearman rank correlation coefficient is $R_{\rm S}=0.44$, indicating that [Fe/H]$_{\rm{knee}}$ is on average higher in galaxies with greater $M_\star$, consistent with the observed trend (e.g. \citealp{tolstoy-09}) for the LGDs.
This correlation is also in qualitative agreement with the $M_V$-[Fe/H]$_{\rm{knee}}$ relation presented by \cite{hendricks-knee} for seven (Sagittarius dSph, Fornax, Sculptor, Ursa Minor, Carina, Draco and Hercules) of the LGDs.  
It is also noteworthy that the simulated MKR is characterised by strong scatter in [Fe/H]$_{\rm{knee}}$ at fixed $M_\star$, in agreement with  observational evidence showing that some of the LGDs exhibit significantly low knee metallicities for their stellar masses (e.g., Fornax \& the UMi dSph; \citealp{hendricks-knee}, and the Magellanic Clouds; \citealp{nidever-lazy-giants}).



\subsubsection{The origin of scatter in [Fe/H]$_{\rm knee}$ at fixed stellar mass}\label{sec4_1_sfhs}

Symbols in Fig.~\ref{fig_mkr_rs} are colour-coded by the residuals for each galaxy about the median relation between $M_{\star,{\rm{plateau}}}$ and $M_{\star}$, calculated non-parametrically using local-weighted scatterplot smoothing (LOWESS; \citealp{lowess-orig}). This information illuminates the origin of the scatter about the MKR (Fig.~\ref{fig_mkr_rs}. The $M_{\star,{\rm{plateau}}}-M_\star$ correlation is unsurprisingly very strong ($R_{\rm S}$=0.96), with a typical scatter of $\simeq 0.24~\rm{dex}$ at fixed $M_\star$. The value for the residual indicates whether a galaxy has a high ($\Delta\log_{10}M_{\star,{\rm plateau}} > 0$) or low ($\Delta\log_{10}M_{\star,{\rm plateau}} < 0$) plateau mass relative to its similarly-massive counterparts. 

\begin{figure}
	\includegraphics[width=\columnwidth]{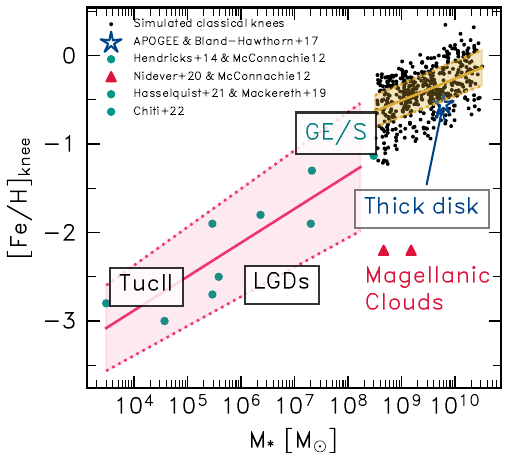}
    \caption{Scatter plot of [Fe/H]$_{\rm{knee}}$ as a function of stellar mass for the simulated galaxies that exhibit a classical knee distribution, in addition to observationally inferred measurements of Local Group dwarf galaxies. It shows that the slope of the simulated MKR corresponds closely to that of the observational data.  The simulation also exhibits a scatter in [Fe/H]$_{\rm{knee}}$ at fixed $M_\star$ in agreement with the observations. In the exceptional cases of the Magellanic Clouds, which exhibit strongly non-monotonic behaviour on the $\upalpha$-Fe plane, it is known that they underwent recent starbursts associated with their infalling onto the MW halo and thus we cannot truly consider them to belong to our category of classical knees.}\label{fig_mkr_comparison}
\end{figure}

The vertical stratification of the symbol colours at fixed $M_\star$ is indicative of a strong positive correlation between [Fe/H]$_{\rm knee}$ and $M_{\star,{\rm{plateau}}}$. The lower panel quantifies this correlation, by showing the Spearman rank correlation coefficient, $R_{\rm S}$, computed for sub-samples within moving bins of $M_\star$ \citep[prior applications of this method can be found in the studies of][]{davies-lowess_1, davies_lowess_2}. A strong, positive correlation is returned with high significance ($p\ll 0.01$) at all $M_\star$. Galaxies with a high knee metallicity for their stellar mass therefore tend to have a higher plateau mass. Whilst this correlation alone does not denote a causal relation between the two, one can reasonably conclude that the relationship is causal based on Fig.~\ref{fig_4panel_singles_knees}, from which we established that the $\upalpha$~knee forms in response to the decline of the SFR, so that the plateau comprises stars formed prior to the SFR peak and the ensuing $\upalpha$~knee formation. Consequently, the Fe synthesised prior to the formation of the $\upalpha$~knee was synthesised by stellar populations comprising the plateau, so it is unsurprising that $\rm{[Fe/H]}_{knee}$ correlates more strongly with $M_{\star,\rm{{plateau}}}$ ($R_{\rm S}=0.64$) than with $M_\star$ ($R_{\rm S}=0.44$).

\subsubsection{Comparing the simulated MKR to that of the Local Group dwarf galaxies}\label{subsec_5_1_LGD_MKR}

It is common in the literature for the stellar masses of dwarf galaxies to be ``predicted`` from their metallicity distribution functions (MDFs; typically represented by the distribution of [Fe/H]), calibrated against the mean MZR (for example that measured by \citealp{kirby13}; e.g. \citealp{carrillo_23}).  
Measurement of the true mean metallicity, $Z$, of a resolved galaxy requires measurements for an unbiased and representative sample of stars, which is at best very costly observationally. 
For any galaxy where the presence of a plateau, $\upalpha$~knee and subsequent shin has been observed in chemical data, the MKR could be used in principle to estimate the total stellar mass of a galaxy, although there is currently a significant lack of measurements for galaxies with mass $\leq10^8~\rm{M}_{\odot}$, and the scatter in the MKR makes this exercise uncertain. 

In Fig.~\ref{fig_mkr_comparison} we display on the MKR plane the 30~kpc aperture stellar masses and fitted values of [Fe/H]$_{\rm{knee}}$ for our classical knee sample and observational data for LGDs (including the Magellanic Clouds) as well as the debris of the Gaia-Enceladus/Sausage accreted system, and the Milky Way high-$\upalpha$ disk.  We use the LGD mass measurements presented by \cite{mcconachie-12}, the thick disk mass estimate of \cite{blandhawthorn-16} and the Gaia-Enceladus/Sausage (GE/S) mass estimate of \cite{mackereth-accreted}.  For the low-mass LGDs we adopt the $\upalpha$~knee metallicities of \cite{hendricks-knee}, for the Magellanic Clouds (MCs) we use the $\upalpha$~knee metallicity estimated by \cite{nidever-lazy-giants} while for GE/S and the Milky Way thick disk we apply the method outlined in \ref{sec2_3_method} on the data from APOGEE.  The data for GE/S stars come from \cite{hasselquist-dwarfs} and the thick disk sample originates from Kisku et al.  (in prep).

Fig.~\ref{fig_mkr_comparison} shows that the slopes of the simulated and Local Group MKRs agree with one another, showing a monotonically increasing trend over the entire mass range sampled.  
A simple linear regression on each data set yields respective slopes of $0.246\pm0.077~\rm{dex}\log_{10}(M_\odot)^{-1}$ and $0.385\pm0.239~\rm{dex}\log{10}(M_\odot)^{-1}$ with the observed relation being very uncertain due to the limited number of data points. We omit the Magellanic Clouds on the basis that even a cursory look at the data \citep[for example, see][]{nidever-lazy-giants, nidever-smash, hasselquist-dwarfs} reveals that they fall under our category of `inverted knees'.  Although they may exhibit a classical knee at the metal-poor end, the bulk of their populations on the $\upalpha$-Fe plane are characterised by a declining-$\upalpha$ plateau at low [Fe/H], turning over to an increasing $\upalpha$-shin, likely caused by a recent burst of star formation.  

    

\section{A comparison of simulated galaxy abundances with those predicted by a GCE}\label{sec6_chemev}

Our explanation of the origin of the knee in $\upalpha$-Fe space differs from the orthodox interpretation. The latter is primarily motivated by the outcomes of galactic chemical evolution (GCE) models, which adopt a broadly similar set of prescriptions for the behaviour with cosmic time of gas inflow and outflow rates, and the gas consumption timescale. The differing physical origin of $\upalpha$-knee formation that one infers from analysis of the simulation and GCE models therefore warrants further investigation.

We argued in $\S$\ref{sec3_results} that the star formation history is a distinguishing characteristic of galaxies that exhibit classical knees in the simulation. 
We therefore examine the element abundance evolution inferred by the \textsc{vice} one-zone GCE model \citep{vice1,vice2,vice3} for a fixed star formation history, which we choose to be that of an exemplar classical knee galaxy ($M_{\star}=10^{9.60}~{\rm{M}_\odot}$) identified in the simulation. 
We compare the outcomes using i) the constant gas consumption timescale, $t_{\rm{g}}=1$~Gyr, assumed by default by the GCE, and ii) an evolving gas consumption timescale similar to that exhibited by the simulated galaxy, which we find to be significantly longer at early cosmic epochs than the default gas consumption timescale assumed by \textsc{vice}. 

\begin{figure}
	\includegraphics[width=\columnwidth]{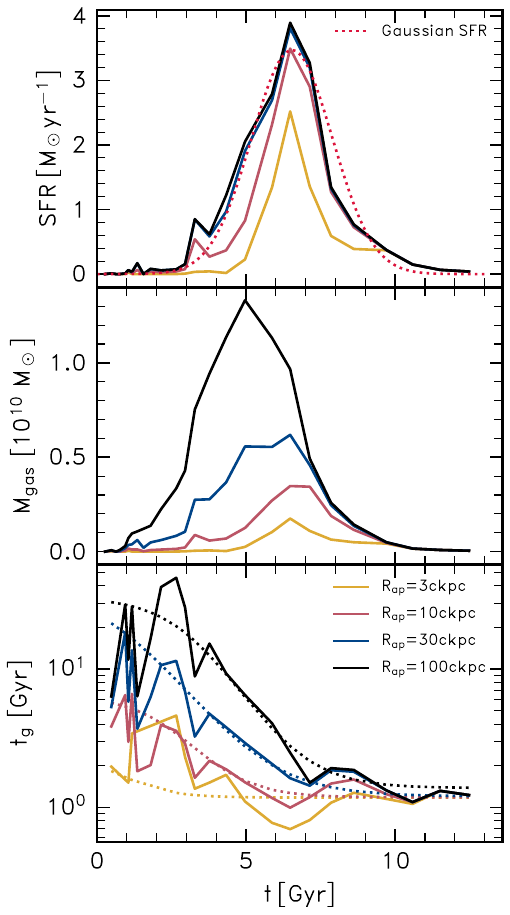}
    \caption{The temporal evolution of the star formation rate (top panel), gas reservoir mass (middle) and gas reservoir consumption timescale (bottom) of the exemplar simulated galaxy with a classical knee. We measure the quantities within spherical apertures of commoving radii $R_{\rm ap}=[3,10,30,100]$ckpc, centred about the minimum of the potential of the galaxy's main progenitor. Analytic fits to the SFR and consumption timescale are shown with dashed lines.}
    \label{fig:tg_merger_tree}
\end{figure}

\subsection{A brief description of the one-zone GCE framework}\label{sec_gce_desc}  

There is an increasing number of chemical evolution frameworks being made publicly available such as the Flexible Chemical Evolution model (FlexCE; \citealp{andrews}), ChemPy (\citealp{chempy}) and the Versatile Integrator for Chemical Evolution (\textsc{vice}; \citealp{vice1}, \citealp{vice2}, \citealp{vice3}). 
Starting from a set of assumed stellar evolution timescales, chemical yields, and an IMF, these one-zone GCE models\footnote{Multi-zone models that predict the radial dependence of element abundance distributions have been constructed, primarily to take advantage of the radially-resolved abundances available for the Milky Way. A comprehensive comparison of numerical simulations and multi-zone models is beyond the scope of this study.} solve a set of differential equations to predict the evolution of element abundances within a galaxy's gas reservoir. They are typically used to infer galaxy formation histories from the distributions of stellar populations of the Milky Way and other Local Group galaxies in various chemical planes.  As indicated by their name, `one-zone' GCE models assume a uniform gas reservoir, all of which is star forming, and into which the nucleosynthetic products of stellar evolution are perfectly mixed upon their release. 

\begin{table}
 \caption{Table of best-fitting model parameters of sigmoid function to merger tree evolutionary tracks of $t_{\rm{g}}$.}
 \label{tab:sigmoid_params}
 \begin{tabular}{lcccr}
  \hline
  $R_{\rm ap}$ [ckpc] & $L$ [Gyr] & $t_0$ [Gyr] & $k$ & $t_{\rm g,0}$ [Gyr] \\
  \hline
  3 & 1.39 & 0.37 & -1.13 & 1.18\\
  10 & 5.21 & 2.37 & -0.99 & 1.18\\
  30 & 38.3 & 0.64 & -0.73 & 1.20\\
  100 & 31.8 & 3.10 & -0.91 & 1.38\\
  \hline
 \end{tabular}
\end{table}

Here we elect to use the \textsc{vice} model. Therein, the SFR is specified by the product of the mass of the (proto-) galaxy's gas reservoir, and the star formation efficiency, where the latter is the inverse of the gas consumption timescale, $t_{\rm g}$. The gas reservoir mass follows from the history of gas inflow, the star formation rate, and the gas outflow rate due to feedback \citep[e.g.,][]{nidever-lazy-giants}. The reduction in gas mass due to outflows is generally specified by the product of the star formation rate and a mass loading factor, $\eta$, i.e. $\dot{M}_{\rm out}(t) = \eta \dot{M}_\star(t)$, usually assumed to be a constant. GCEs also typically adopt fixed values of the gas consumption timescale, of $t_{\rm g}\simeq 1\,{\rm Gyr}$ \citep[e.g.][]{weinberg-sudden-events}, though there are exceptions in the literature \citep[e.g.,][]{hasselquist-dwarfs,vasini_lmc_23}. This timescale is typically motivated from the observationally-inferred consumption timescale of molecular gas in nearby galaxies \citep[$\simeq 0.5-2\,{\rm Gyr}$, e.g.][]{leroy08,leroy_13,sun_23}, with observations of distant galaxies possibly indicating a shorter molecular gas consumption 
timescale \citep{tacconi_18}.

\subsection{The evolution of an exemplar simulated galaxy with a classical knee}\label{sec_eagle_gce}

The evolution of the gas reservoir's mass, $M_{\rm gas}(t)$, and parameters such as $t_{\rm g}$ and $\eta$, clearly have a marked influence on the resulting chemical evolution predicted by GCEs \citep[e.g.][]{andrews, vice1}. 
Since in \textsc{vice} the star formation rate is specified by the ratio of the gas mass and the gas consumption timescale, our choice of adopting a fixed star formation history means that the gas inflow history is specified by the gas consumption timescale, i.e. $M_{\rm gas}(t) = \dot{M}_\star(t)~t_{\rm g}$, where $t_{\rm g}$ can be fixed or time evolving (by default \textsc{vice} assumes a fixed gas consumption timescale of $t_{\rm g}=1\,{\rm Gyr}$). We therefore begin by examining the evolving gas reservoir mass, and its corresponding consumption timescale, for our exemplar simulated galaxy. We focus here on $M_{\rm gas}$ and  $t_{\rm g}$ and omit $\eta$, as the latter has a less significant impact on the predicted element abundances in \textsc{vice}. We direct readers interested in the outflow mass loadings of EAGLE galaxies to the study of \citet{mitchell_20}, who showed that (at a fixed redshift and for a fixed aperture) mass loading increases with decreasing galaxy mass. We note therefore that examination of a time- or galaxy mass-dependent outflow mass loading in \textsc{vice} would be an interesting avenue of inquiry, however this is beyond the scope of our investigation.

Fig.~\ref{fig:tg_merger_tree} shows the time evolution of the star formation rate, $\dot{M}_\star$ (top panel), the gas reservoir mass, $M_{\rm gas}$ (middle panel), and the consumption timescale of the reservoir, $t_{\rm g}=M_{\rm gas}/\dot{M}_\star$ (bottom panel). $M_{\rm gas}$ and $\dot{M}_\star$ are measured by summing the corresponding quantities from particles enclosed within spherical apertures of comoving radii $R_{\rm ap}=[3,10,30,100]$~ckpc, centred about the minimum of the potential of the galaxy's main progenitor. We identify the latter using the simple merger tree scheme described by \citet{pfeffer-emosaics}. The aperture measurements are shown using curves coloured yellow, red, blue and black, respectively. We measure the gas mass\footnote{We consider all gas, irrespective of its ionisation state, since the nucleosynthetic products of stellar evolution are not preferentially mixed with gas of any particular state.} and compute its associated consumption timescale within multiple apertures to obtain a sense of the uncertainty on these measurements when comparing with the predictions from the GCE. This is because it is not obvious how to most meaningfully compare the uniform gas reservoir of a one-zone GCE with the multi-phase gas distributions of hydrodynamically-simulated galaxies, within which only a fraction of the gas will be star forming.  We stress that one should not consider only those particles that are star forming, because this would underestimate the mass of the reservoir into which newly-synthesised heavy elements are mixed. 

\begin{figure}
\includegraphics{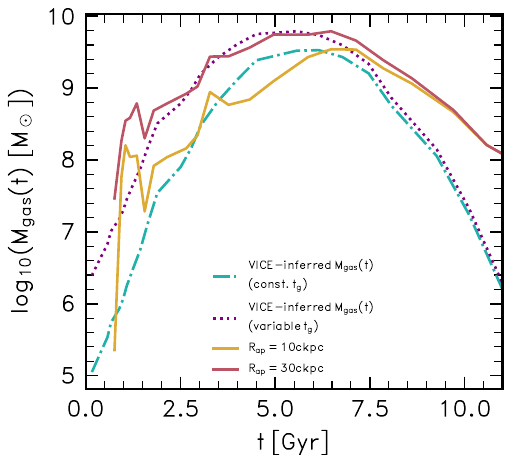}
    \caption{The evolution of the gas reservoir mass inferred by the \textsc{vice} GCE, for the adopted star formation history, in the cases of a constant gas consumption timescale ($t_{\rm g}=1$~Gyr, cyan dot-dashed curve) and a variable $t_{\rm g}$ similar to that exhibited by gas within $30\,{\rm ckpc}$ of the main progenitor of the example simulated galaxy (dotted purple line).  The true gas mass of the simulated galaxy, measured within spheres of radius $10\,{\rm ckpc}$ and $30\,{\rm ckpc}$ are shown with solid yellow and red curves, respectively.}\label{fig:mgas_comparison}
\end{figure}

The upper panel reveals that, consistent with the interpretation of Fig.~\ref{fig_4panel_singles_knees}, the exemplar simulated galaxy exhibits a peaked star formation history, which reaches a maximum $\dot{M}_\star \simeq 4\,{\rm M}_\odot\,{\rm yr}^{-1}$ at $t\simeq 6\,{\rm Gyr}$. 
The aperture measurements highlight that the star formation is centrally concentrated (i.e. entirely enclosed within $R_{\rm ap}=30\,{\rm ckpc}$, with most enclosed within $10 \,{\rm ckpc}$ ). 
The middle panel shows the evolving mass of gas within the apertures centred on the main progenitor, which broadly follow the same behaviour as the SFR, though for the $R_{\rm ap} = 100\,{\rm ckpc}$ aperture the peak gas mass of $2\times 10^{10}\,{\rm M}_\odot$ is reached at $t\simeq 4\,{\rm Gyr}$, giving a sense of the time taken for inflows onto the halo to reach the galaxy. 
In the bottom panel we show the consumption timescale of the gas reservoir, by taking the ratio of the gas mass and star formation rate measured within each of the apertures. It is immediately apparent that at early times ($t\lesssim 5\,{\rm Gyr}$), the gas consumption timescale of the simulated galaxy is significantly longer than the default of $1\,{\rm Gyr}$ assumed by \textsc{vice}. 
For the case in which we consider the gas within $10\,{\rm ckpc}$, which can be considered conservative since this aperture only just envelops all of the star-forming gas (and thus plausibly excludes cold gas into which newly-synthesised heavy elements will be mixed), we recover a gas consumption timescale $t_{\rm g}>10\,{\rm Gyr}$ for $t\lesssim 2\,{\rm Gyr}$. 
For a fixed star formation history, a longer gas consumption timescale implies the presence of a larger (and less readily enriched) gas reservoir so, as we explore in more detail in \S\ref{sec:one_gce_model},  the implications for the evolution of element abundances predicted by GCEs are significant. 

To simplify the use of the simulated galaxy's histories of star formation and gas consumption timescale as inputs to \textsc{vice}, we approximate them with analytic functions. We use a Gaussian function for the star formation history,
\begin{equation}
    \label{eq_gaussian_sfr}
    \dot{M}_\star(t) = A_0\exp{\left(\frac{t-t_0}{\sigma}\right)^2},
\end{equation}
\noindent and recover best-fit parameters to the SFR within $R_{\rm ap}=30\,{\rm ckpc}$ of $A_0=3.47~{\rm M_\odot~yr^{-1}}$, $t_0=6.54$~Gyr and $\sigma=1.92$~Gyr. For the gas consumption timescale we fit a sigmoid function:
\begin{equation}
    t_{\rm g}(t) = t_{\rm g,0} + \frac{L}{1+\exp{[-k(t-t_0)}]}. \label{eq_tg_model}
\end{equation}
whose best-fit parameters for the four apertures are specified in Table \ref{tab:sigmoid_params}. The best fits are shown with dashed lines in the top and bottom panels of \ref{fig:tg_merger_tree}.



\subsection{A one-zone GCE model based on the SFH of a simulated galaxy with a classical knee}
\label{sec:one_gce_model}

We construct a \textsc{vice} model using the parametrised SFH of a realistic simulated galaxy with a classical knee (i.e. equation \ref{eq_gaussian_sfr}) as an input. We mostly adopt the default {\textsc{vice}} parameters, an exponential SN~Ia delay-time distribution, with a minimum delay time $t_{\rm D,min}=0.15~{\rm{Gyr}}$ and e-folding timescale $\tau_{\rm Ia}=1.5~{\rm{Gyr}}$. We also assume a \cite{chabrier_IMF} IMF with lower and upper mass limits $0.08~{\rm{M}_\odot}$ and $100~{\rm{M}_\odot}$ respectively.


Using the SN~II ejected masses reported in Tables 7 and 10 of \cite{portinari-yields}, we compute the IMF-averaged yields from SN~II for O and Fe by numerical integration using Equation 6 from \cite{vice1}:
\begin{equation}
    y_{\rm x}^{\rm cc}=\frac{\int_{\rm M_{SN}}^{\rm u}m_{\rm x}\frac{{\rm d}N}{{\rm d}m}{\rm d}{m}}{\int_{\rm l}^{\rm u} m \frac{{\rm{d}}N}{{\rm{d}}m} {\rm d}m},
\end{equation}
where $m_x$ is the mass of element $x$ ejected in the explosion of a star of initial mass $m$ and $\frac{{\rm d}N}{{\rm d}M}$ is the assumed stellar IMF for which we adopt the \cite{chabrier_IMF} IMF with a lower stellar mass limit $l=0.08~{\rm M_{\odot}}$ and an upper stellar mass limit $u=100~{\rm{M}_\odot}$. $\rm{M}_{SN}$ is the minimum initial mass of a SN~II progenitor, assumed to be 6 $\rm{M}_\odot$ (\citealp{portinari-yields} include electron capture SN, whose progenitors are assumed to be quasi-massive stars with initial masses $M_{\rm i,~prog}\approx6-7~{\rm M_\odot}$). By taking the mean yield between metallicities $Z=0.004$ and $Z=0.05$, we derive net IMF-averaged yields for the \cite{portinari-yields} yield tables of $\rm{}y_{II}^{O}=0.0172$ and $\rm{}y_{II}^{Fe}=0.00148$ for O and Fe, respectively. The EAGLE model rescales the SN~II yield of Fe by a factor of 0.5, and we adopt the same SN~II yield for Fe when running \textsc{VICE}. These yields place the pure SN~II plateau at $\rm [O/Fe]=0.67$, in agreement with the low-metallicity trends seen in the simulations (see Fig.~\ref{fig_threepanel_afe}). The Fe yield from SN~Ia is computed simply by taking:
\begin{equation}
    y_{\rm Ia}^{\rm Fe} = \nu_{\rm Ia} \times m_{\rm Fe},
\end{equation}
where $\nu_{\rm Ia}$ is the total number of SN~Ia per unit initial stellar mass, equal to $\nu=2\times10^{-3}~{\rm M_{\odot}}^{-1}$. Adopting the SN~Ia yields of \cite{thielemanns_03} (an ejected Fe mass per SN Ia of $m_{\rm ej,Fe~Ia}=0.769~{\rm M_\odot}$), we derive $y_{\rm Ia}^{\rm Fe}=0.00154$.

\begin{figure*}
    \includegraphics{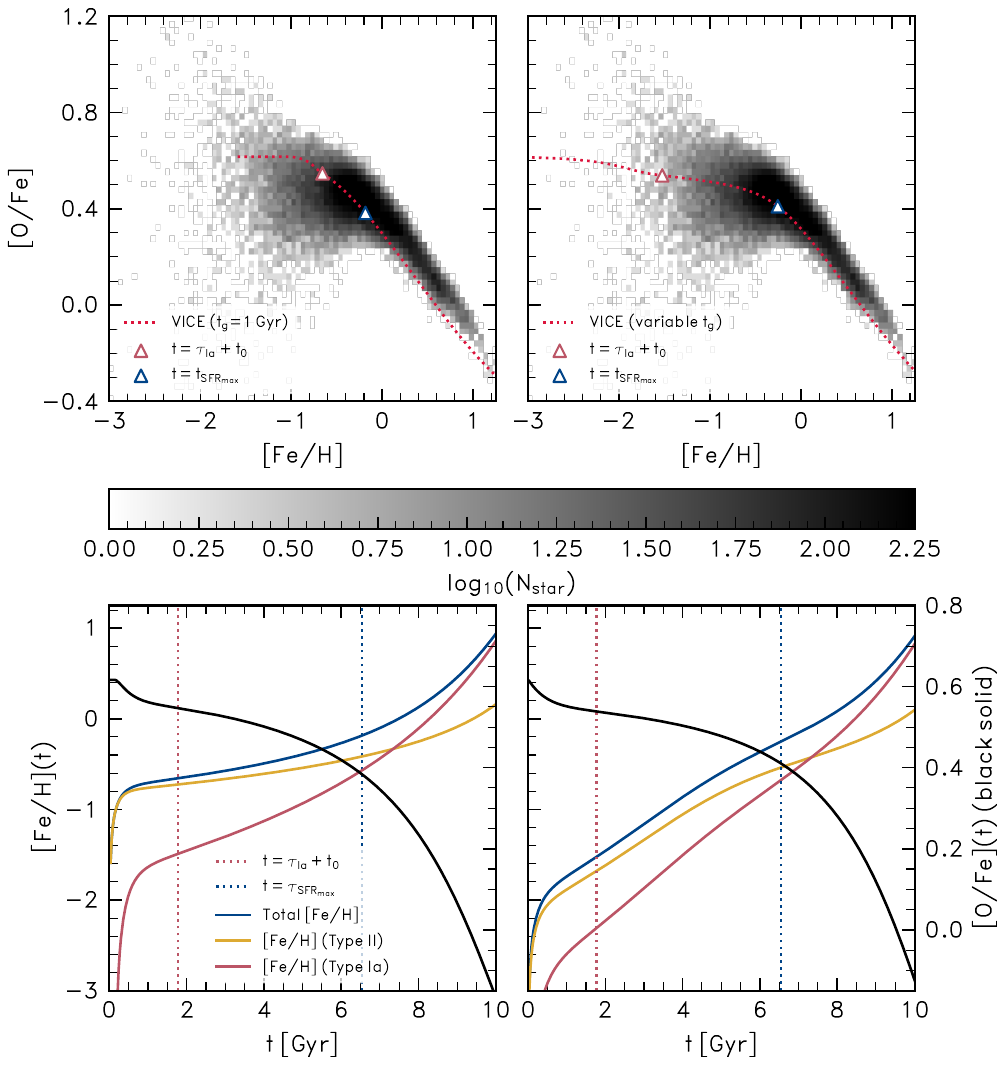}
    \caption{The element abundance evolution predicted by the \textsc{vice} GCE when given, as an input, the SFH of an exemplar simulated galaxy with a classical knee. The top and bottom rows correspond to the constant $t_{\rm g}$ and variable $t_{\rm g}$ cases, respectively. The left column shows the temporal evolution of the gas reservoir's path through the $\upalpha$-Fe plane (yellow dotted curve). Red overlaid triangles denote the time at which one e-folding timescale of the assumed DTD ($\tau_{\rm{Ia}}=1.5~{\rm Gyr}$)has elapsed after the formation of the first stellar particle ($t_0$) and blue triangles denote the epoch at which the SFR peaks. The 2-dimensional histogram shows the stellar particle distribution from the exemplar simulated galaxy. The right column shows the temporal evolution of [Fe/H] (left axis, solid blue curve), with yellow and red curves denoting the contributions from SN~II and SN~Ia, respectively, and [O/Fe] (right axis, solid black curve). Vertical dotted lines correspond to the same epochs denoted by the triangles in the left column. The low-mass of the one-zone gas reservoir implied by the constant $t_{\rm g}$ case leads to its rapid enrichment to [Fe/H]$>-1$, so the plateau corresponds to only a brief period in the galaxy's evolution, comprises little mass and is more $\upalpha$-rich than that of the simulated galaxy. In contrast the (initially) more massive reservoir implied by variable $t_{\rm g}$ case is more gradually enriched, resulting in the later formation of the knee at a lower [O/Fe], similar to that of the simulated galaxy.}
    \label{fig:chemev_comparison}
\end{figure*}

The impact of adjusting the assumed gas consumption timescale for a fixed star formation history in the GCE is illustrated by Fig.~\ref{fig:mgas_comparison}: the mass of the one-zone gas reservoir inferred by \textsc{vice} for the $t_{\rm g}=1\,{\rm Gyr}$ is denoted by the cyan dot-dashed curve, and that inferred from the fit to the gas consumption timescale of the gas within $30\,{\rm ckpc}$ of the simulated galaxy's main progenitor is shown with a purple dotted curve. 
At early epochs ($t\lesssim 2.5\,{\rm Gyr}$), the inferred gas reservoir is a factor of $\sim 10$ more massive in the latter case, with the offset diminishing at later times as the fit to the simulated galaxy's gas consumption timescale declines and, for $t\gtrsim 7\,{\rm Gyr}$, converges towards a value similar to the default assumption of $t_{\rm g} = 1\,{\rm Gyr}$. 
The solid yellow and red curves, repeated from the centre panel of Fig.~\ref{fig:tg_merger_tree}, denote the total gas mass of the simulated galaxy's main progenitor within spheres of radius $10\,{\rm ckpc}$ and $30\,{\rm ckpc}$, respectively. 
Note that the GCE gas reservoir mass based on the simulated galaxy's gas consumption timescale (dotted red curve) need not closely trace the simulated galaxy's actual gas mass within the same aperture (solid red curve): the GCE model reservoir is uniformly star-forming, whereas only some of the simulated galaxy's associated gas has density above the adopted density threshold for star formation.  In addition, there is a mild dependence of the star formation rate on the gas pressure $\dot{m}_\star \propto P^{1/5}$ \citep{schaye-ks-law}, which is unaccounted for in this procedure and contributes to the mismatch.

Fig.~\ref{fig:chemev_comparison} shows the resulting evolutionary tracks of [Fe/H](t) and [O/Fe](t), in two different chemical planes. The left panel shows the $\upalpha$-Fe plane, while the right panel shows the evolutionary tracks as a function of cosmic time for these abundance ratios, split into the contributions of SN~II and SN~Ia, respectively. Red triangles on the evolutionary tracks indicate the abundances at $t=\tau_{\rm Ia}+t_0$, where $\tau_{\rm Ia}$ is the e-folding timescale of the assumed SN~Ia DTD and $t_0$ is the formation time of the first stellar particle; blue triangles indicate the time at which the star formation rate peaks and the knee forms in the simulation ($\simeq$6.5~Gyr). The same times are indicated by vertical dotted lines in the right panels. The top row corresponds to the fixed $t_{\rm g}=1$ Gyr case, the bottom row to the variable case.

In both cases, the predicted metallicity of the knee closely corresponds with that exhibited by the simulated galaxy (see Fig.~\ref{fig_ck_fit}). 
However, the constant $t_{\rm g}$ case produces a plateau at [O/Fe]$\simeq$0.6 and a knee forms at $t \simeq \tau_{\rm Ia}$, preceding the time at which the knee forms in the simulated galaxy by $\simeq 5\,{\rm Gyr}$. 
The plateau therefore corresponds to only a brief and early period of the galaxy's evolution, during which there is little enrichment by SN~Ia. In contrast, the (initially) more massive gas reservoir implied by the variable $t_{\rm g}$ case is more gradually enriched, leading to the formation of a gently declining high-$\upalpha$ plateau over a prolonged period, so that the knee forms at a much later time, broadly corresponding to that at which the star formation rate begins to decline. 
It is noteworthy that the two cases, implementing identical yields for SN~II and SN~Ia and the same IMF, yield plateaus of significantly different $\alpha$ richness. The constant $t_{\rm g}$ case features a short-lived plateau at the [O/Fe] value corresponding to the IMF-averaged yield of SN~II, indicating that the natal gas of these stars was enriched exclusively by SN~II. 
The same feature is seen in in the variable $t_{\rm g}$ case for the very low metallicity ([Fe/H]$\lesssim -2.5$) part of the plateau. However, at greater metallicity (corresponding to $t \gtrsim \tau_{\rm Ia}$), [O/Fe] declines steadily, which is indicative of a growing contribution to enrichment by SN~Ia.

The evolution of stellar abundances shown in the right panels of Fig.~\ref{fig:chemev_comparison}, analysed in the light of Fig.~\ref{fig:mgas_comparison}, highlights why, for a fixed star formation history, the assumed gas consumption timescale has such a marked influence on the evolution of the element abundances. 
The much less massive gas reservoir in the constant $t_{\rm g}$ case is quickly enriched, reaching [Fe/H]$ =-1$ at $t\simeq\tau_{\rm Ia}$, whereas the more massive reservoir implied by the variable $t_{\rm g}$ case is enriched gradually, attaining [Fe/H]$ =-1$ at $t\simeq5$~Gyr.  This has significant implications for the resulting stellar metallicity distribution function (MDF), which we show in Fig.~\ref{fig:mdfs}. Since the star formation rate adopted is relatively low at early times, the brief window of time before the gas reservoir is enriched in the $t_{\rm g}=1$~Gyr case leads to a paucity of metal poor stars (solid blue curve), compared to the MDF of the simulated galaxy (red histogram), which in turn is broadly reproduced by the variable $t_{\rm g}$ case (solid yellow curve). 

Our experiments with \textsc{vice} demonstrate that the nucleosynthetic products of the stellar populations of the progenitors of Milky Way-like galaxies are almost certainly mixed into a reservoir of gas that is significantly more massive and spatially-extended than the nascent galaxy's reservoir of molecular gas, with much of this larger reservoir either forming stars at a later time, or being expelled from the circumgalactic environment by feedback-driven outflows. Approximating this scenario with a one-zone GCE model therefore requires the adoption of a longer consumption timescale than that one infers from observations of molecular gas in distant galaxies \citep[e.g.][]{tacconi_18}, ensuring that newly-synthesised metals are sufficiently diluted.

\subsection{The nature of the [\texorpdfstring{$\upalpha$}{alpha}/Fe] plateau}

Both of the \textsc{vice} models exhibit a nearly flat $\ofe$ trend between 
$\feh = -1.5$ and $-0.5$, but the nature of this trend is quite different
in the two cases.  For the constant-$t_{\rm g}$ model, the plateau in this regime
still reflects the Type II supernova yield ratio.  For the variable-$t_{\rm g}$
model, the $\ofe$ in this regime is about 0.1-dex below this ratio, and
it stays roughly flat because the rising SFR allows current Type II enrichment
to keep pace with the growing Type Ia enrichment from the growing evolved
population.  The knee arises when the SFR begins to decline, but it would
be present, albeit less sharply, even if the SFR merely flattened and 
became constant.

The variable-$t_{\rm g}$ model is reminiscent of the model of early Milky Way
chemical evolution proposed by Conroy et al. (\citeyear{conroy_h3_23}; see
\cite{chen_23} for a broader sample of similar models), which is motivated
by the observed $\afe-\feh$ trends in the H3 survey of the stellar halo.
There the star formation efficiency grows rapidly from a low value of
$t_{\rm g}^{-1} = (50~\Gyr)^{-1}$ at $t<2.5~\Gyr$ to $(2.36~\Gyr)^{-1}$ at
$t>3.7~\Gyr$.  The $\afe$ track declines from +0.6 at low $\feh$ to about
+0.25 at $\feh=-1.3$, then rises gently in response to the accelerating
SFR before turning downward again above $\feh \approx -0.5$.
The behaviour of our variable-$t_{\rm g}$ model is milder than this, with
$\ofe$ always declining, but it is similar enough to suggest that the kind
of evolution envisioned in the \cite{conroy_h3_23} model arises fairly commonly
in EAGLE galaxies.  This behaviour also supports the contention of 
\cite{Maoz-Graur-Alpha} that the $\afe$ plateau observed in the Galactic thick disk
already includes substantial Type Ia supernova enrichment.

\begin{figure}
	\includegraphics[width=\columnwidth]{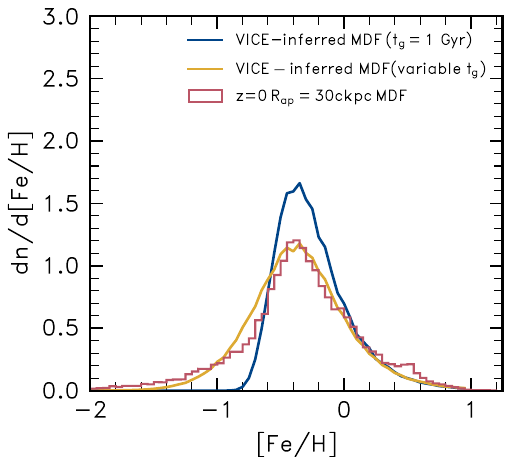}
    \caption{Metallicity distribution function (MDF), using [Fe/H] as a proxy for metallicity, of the exemplar simulated galaxy with a classical knee (red histogram), and of the VICE GCE model using the simulated galaxy's SFH as an input, for two cases of the adopted gas reservoir consumption timescale. The blue curve corresponds to the VICE default, $t_{\rm g}=1~{\rm Gyr}$, the yellow curve corresponds to a variable gas consumption timescale similar to that exhibited by the simulated galaxy (see Eq.~\ref{eq_tg_model} VICE reproduced the simulated galaxy's MDF closely in the latter case but the assumption of a constant, short gas consumption time results in a paucity of low metallicity ([Fe/H]$\lesssim -1$) stars. }
    \label{fig:mdfs}
\end{figure}

As Fig.~\ref{fig:mdfs} shows, the MDF is a good diagnostic for distinguishing
these two classes of GCE models.  The age-metallicity relation of metal-poor
stars is also a valuable diagnostic for the early behaviour of $t_{\rm g}$,
especially if ages are accurate enough to resolve the evolution from
$\feh \approx -1.5$ to -0.5 in a stellar population that can reasonably
be ascribed to a single well-mixed progenitor.

We have examined here the star formation history of a single exemplar galaxy, but we note that in one-zone GCE models where the star formation rate is set by the ratio of the gas reservoir mass and the gas consumption timescale, the assumption of a short gas consumption timescale will result in rapid enrichment of the gas reservoir for any star formation rate. As such, we argue that the assumption of a gas consumption timescale that is significantly shorter at early epochs than is indicated by realistic cosmological hydrodynamical simulations of galaxies, leads to the unrealistic prediction that the natal gas of stars comprising $\upalpha$-rich plateaus was enriched exclusively by SN~II.


\section{Summary of results and conclusions}\label{sec5_summary}

We have examined the stellar [$\upalpha$/Fe]-[Fe/H] distribution of $\simeq1000$ galaxies of present-day mass $M_\star \gtrsim  10^{8.5}~{\rm{M}_{\odot}}$ that form in a high-resolution simulation from the EAGLE suite. The simulation has been shown elsewhere to reproduce a wide range of key properties of the observed galaxy population. Our main conclusions can be summarised as follows:

\begin{enumerate}
    \item The simulated galaxies exhibit a diversity of morphologies in the $\upalpha$-Fe plane (Fig.~\ref{fig_threepanel_afe}). The distributions can be broadly categorised into three types: i) ``classical knees'' which exhibit a high-$\upalpha$ plateau followed by a declining-$\upalpha$ shin; ii) ``single-slopes'' which have mono-gradient $\upalpha$-Fe planes, and; iii) more complicated distributions such as ``inverted knees'', where there is a sequence of increasing $\upalpha$ at high [Fe/H].
    
    \item The galaxy population is dominated by classical knees, but there is a significant mass dependence to the relative population of each type (Fig.~\ref{fig_frac_gals}). Single slopes are more common for $M_\star \lesssim 10^9\,{\rm M}_\star$, but their frequency decreases monotonically with $M_\star$, while that of classical knees increases. We find that the slope of the plateau on the $\upalpha$-Fe plane is consistent with that of the single slopes but that of the shin is typically steeper by a factor $\simeq2$ (Fig.~\ref{fig_gal_slopes}). This indicates that the relative contributions to enrichment by SN~II and SN~Ia is similar in plateaus and single slope galaxies, whereas the shin is characterised by a diminished contribution from SN~II.

    \item Galaxies with classical knees exhibit peaked SFHs, typically reaching their maximum at $t\simeq 5-7\,\rm{Gyr}$ ($z \simeq 1.2-0.7$)  before exhibiting a sustained decline until the present day. 
    Contrary to the predictions of GCE models, we find that the typical time of $\upalpha$~knee formation, $t_{\rm knee}$, does not coincide with the time at which SN~Ia first contribute significantly to the chemical enrichment (Fig.~\ref{fig_4panel_singles_knees}).  Rather, $t_{\rm knee}$ corresponds roughly to the time at which the SFR begins to decline.  The $\upalpha$-poor, Fe-rich shin forms when the star formation rate declines, so that SN~II do not enrich the ISM as effectively as SN~Ia. This results in the slope of the $\upalpha$-Fe shin being significantly steeper than that of the high-$\upalpha$ plateau as enrichment switches from being dominated by young stellar populations to old stellar populations.

    \item Conversely, the SFHs of galaxies showing single-slope distributions in the $\upalpha$-Fe plane are not as strongly peaked as those with classical knees.  As a result, the stars of single slopes galaxies have a broader spread of stellar ages than those of classical knees.

    \item Simulated galaxies presenting classical knees exhibit a correlation between the metallicity of the $\upalpha$~knee and galaxy mass, albeit with significant scatter in [Fe/H]$_{\rm knee}$ at fixed $M_\star$ (Fig.~\ref{fig_mkr_rs}). The residuals about this scaling relation (which we dub the MKR) correlate strongly with the fraction of the mass in the high-$\upalpha$ plateau.

    \item The MKR exhibited by the simulated galaxies has slope $0.246\pm0.077~\rm{dex}\log_{10}(M_\odot)^{-1}$, consistent with the Local Group dwarf galaxy MKR which has slope $0.385\pm0.239~\rm{dex}\log{10}(M_\odot)^{-1}$ (Fig.~\ref{fig_mkr_comparison}). The simulations also broadly reproduce the observed scatter in the MKR. \cite{nidever-lazy-giants} reported a strong deviation of the Magellanic Clouds from that relation. However, although it may have a classical knee at very low metallicity, we find that the bulk of their stellar populations do not follow the distribution of classical knees in the $\upalpha$-Fe plane, which may explain the discrepancy.
    

    \item One-zone chemical evolution models, using as an input the SFH of an exemplar realistic simulated galaxy that exhibits a classical knee, do not reproduce the element abundance history of the simulated galaxy when adopting a constant, and relatively short, gas consumption timescale, typically assumed to be $t_{\rm g}\simeq {\rm 1~Gyr}$ in the literature (Fig.~\ref{fig:chemev_comparison}). 
    The mass of the one-zone gas reservoir inferred by GCE at early epochs is much smaller than the gas mass associated with the main progenitor of the simulated galaxy, resulting in the reservoir being enriched to ${\rm [Fe/H]=-1}$ almost instantaneously. This yields a plateau in the $\upalpha$-Fe plane that is exclusively enriched by SN~II, but whose formation is so rapid that it comprises little stellar mass, resulting in a metallicity distribution function that lacks a significant population of metal poor (${\rm [Fe/H]<-1}$) stars. In contrast, by adopting a gas consumption timescale that evolves in a similar fashion to that of the simulated galaxy, the stars of the plateau form over several gigayears and are significantly enriched by both SN~II and SN~Ia.  
\end{enumerate}

Our findings motivate a revision of the canonical interpretation of the origin of the distinct sequences that are often, but not always, observed in the [$\upalpha$/Fe]-[Fe/H] distribution traced by the stellar populations of Local Group galaxies. At the root of our conclusions is a physical formulation of the star formation efficiency.  The latter is a parameter of GCE models that is difficult to constrain, and which we have estimated here from the properties of the gaseous environment of the main progenitor of a realistic simulated galaxy. This application is a modest example of a growing trend in the field of Galactic Archaeology, whereby the development of increasingly sophisticated cosmological numerical simulations, on a par with the ever growing size and complexity of observational maps of the chemodynamical properties of Local Group galaxies, is deepening our understanding of the physical processes that dictate their formation and evolution.






\section*{Acknowledgements}\label{sec7_ack}

ACM is supported by an STFC doctoral studentship within the Liverpool Centre for Doctoral Training for Innovation in Data Intensive Science (LIV.INNO), hosted by
Liverpool John Moores University and the University of Liverpool. RAC was supported by a Royal Society University Research Fellowship during part of this study. JP is supported by the Australian government through the Australian Research Council’s Discovery Projects funding scheme (DP220101863). TT is supported by Science and Technology Facilities Council (STFC) astronomy consolidated grant ST/P000541/1. Analyses and plots presented in this article used {\textsc{IPYTHON}} and packages in the {\textsc{SCIPY}} ecosystem (\citealp{jones_scipy, hunter_matplotlib,perez_ipython,seabold_statsmodels,vanderwalt_numpy}). The study made use of the Prospero high-performance computing facility at Liverpool John Moores University, and the DiRAC Data Centric system at Durham University, operated by the Institute for Computational Cosmology on behalf of the STFC DiRAC HPC Facility (\url{http://www.dirac.ac.uk/}).

\section*{Data Availability}\label{sec8_data}
The data underlying the results from EAGLE in this article will be shared on reasonable request to the corresponding author. All APOGEE DR17 data upon which this study are based are publicly available and can be found at \url{https://www.sdss4.org/dr17/} .
\bibliographystyle{mnras}
\bibliography{refs} 

\begin{thebibliography}{}
\makeatletter
\relax
\def\mn@urlcharsother{\let\do\@makeother \do\$\do\&\do\#\do\^\do\_\do\%\do\~}
\def\mn@doi{\begingroup\mn@urlcharsother \@ifnextchar [ {\mn@doi@}
  {\mn@doi@[]}}
\def\mn@doi@[#1]#2{\def\@tempa{#1}\ifx\@tempa\@empty \href
  {http://dx.doi.org/#2} {doi:#2}\else \href {http://dx.doi.org/#2} {#1}\fi
  \endgroup}
\def\mn@eprint#1#2{\mn@eprint@#1:#2::\@nil}
\def\mn@eprint@arXiv#1{\href {http://arxiv.org/abs/#1} {{\tt arXiv:#1}}}
\def\mn@eprint@dblp#1{\href {http://dblp.uni-trier.de/rec/bibtex/#1.xml}
  {dblp:#1}}
\def\mn@eprint@#1:#2:#3:#4\@nil{\def\@tempa {#1}\def\@tempb {#2}\def\@tempc
  {#3}\ifx \@tempc \@empty \let \@tempc \@tempb \let \@tempb \@tempa \fi \ifx
  \@tempb \@empty \def\@tempb {arXiv}\fi \@ifundefined
  {mn@eprint@\@tempb}{\@tempb:\@tempc}{\expandafter \expandafter \csname
  mn@eprint@\@tempb\endcsname \expandafter{\@tempc}}}

\bibitem[\protect\citeauthoryear{{Andrews}, {Weinberg}, {Sch{\"o}nrich}  \&
  {Johnson}}{{Andrews} et~al.}{2017}]{andrews}
{Andrews} B.~H.,  {Weinberg} D.~H.,  {Sch{\"o}nrich} R.,   {Johnson} J.~A.,
  2017, \mn@doi [\apj] {10.3847/1538-4357/835/2/224}, \href
  {https://ui.adsabs.harvard.edu/abs/2017ApJ...835..224A} {835, 224}

\bibitem[\protect\citeauthoryear{{Asplund}, {Grevesse}, {Sauval}  \&
  {Scott}}{{Asplund} et~al.}{2009}]{asplund-eagle-solar}
{Asplund} M.,  {Grevesse} N.,  {Sauval} A.~J.,   {Scott} P.,  2009, \mn@doi
  [\araa] {10.1146/annurev.astro.46.060407.145222}, \href
  {https://ui.adsabs.harvard.edu/abs/2009ARA&A..47..481A} {47, 481}

\bibitem[\protect\citeauthoryear{{Baldry} et~al.,}{{Baldry}
  et~al.}{2012}]{baldry_gama}
{Baldry} I.~K.,  et~al., 2012, \mn@doi [\mnras]
  {10.1111/j.1365-2966.2012.20340.x}, \href
  {https://ui.adsabs.harvard.edu/abs/2012MNRAS.421..621B} {421, 621}

\bibitem[\protect\citeauthoryear{{Bastian}, {Pfeffer}, {Kruijssen}, {Crain},
  {Trujillo-Gomez}  \& {Reina-Campos}}{{Bastian}
  et~al.}{2020}]{bastian-emosaics}
{Bastian} N.,  {Pfeffer} J.,  {Kruijssen} J.~M.~D.,  {Crain} R.~A.,
  {Trujillo-Gomez} S.,   {Reina-Campos} M.,  2020, \mn@doi [\mnras]
  {10.1093/mnras/staa2453}, \href
  {https://ui.adsabs.harvard.edu/abs/2020MNRAS.498.1050B} {498, 1050}

\bibitem[\protect\citeauthoryear{{Bensby}, {Feltzing}  \& {Oey}}{{Bensby}
  et~al.}{2014}]{Bensby2014}
{Bensby} T.,  {Feltzing} S.,   {Oey} M.~S.,  2014, \mn@doi [\aap]
  {10.1051/0004-6361/201322631}, \href
  {https://ui.adsabs.harvard.edu/abs/2014A&A...562A..71B} {562, A71}

\bibitem[\protect\citeauthoryear{{Bland-Hawthorn} \&
  {Gerhard}}{{Bland-Hawthorn} \& {Gerhard}}{2016}]{blandhawthorn-16}
{Bland-Hawthorn} J.,  {Gerhard} O.,  2016, \mn@doi [\araa]
  {10.1146/annurev-astro-081915-023441}, \href
  {https://ui.adsabs.harvard.edu/abs/2016ARA&A..54..529B} {54, 529}

\bibitem[\protect\citeauthoryear{{Booth} \& {Schaye}}{{Booth} \&
  {Schaye}}{2009}]{booth2009}
{Booth} C.~M.,  {Schaye} J.,  2009, \mn@doi [\mnras]
  {10.1111/j.1365-2966.2009.15043.x}, \href
  {https://ui.adsabs.harvard.edu/abs/2009MNRAS.398...53B} {398, 53}

\bibitem[\protect\citeauthoryear{{Bower}, {Schaye}, {Frenk}, {Theuns},
  {Schaller}, {Crain}  \& {McAlpine}}{{Bower} et~al.}{2017}]{bower17}
{Bower} R.~G.,  {Schaye} J.,  {Frenk} C.~S.,  {Theuns} T.,  {Schaller} M.,
  {Crain} R.~A.,   {McAlpine} S.,  2017, \mn@doi [\mnras]
  {10.1093/mnras/stw2735}, \href
  {https://ui.adsabs.harvard.edu/abs/2017MNRAS.465...32B} {465, 32}

\bibitem[\protect\citeauthoryear{{Carlin}, {Sheffield}, {Cunha}  \&
  {Smith}}{{Carlin} et~al.}{2018}]{carlin-explosive-alpha}
{Carlin} J.~L.,  {Sheffield} A.~A.,  {Cunha} K.,   {Smith} V.~V.,  2018,
  \mn@doi [\apjl] {10.3847/2041-8213/aac3d8}, \href
  {https://ui.adsabs.harvard.edu/abs/2018ApJ...859L..10C} {859, L10}

\bibitem[\protect\citeauthoryear{{Carrillo}, {Deason}, {Fattahi}, {Callingham}
  \& {Grand}}{{Carrillo} et~al.}{2023}]{carrillo_23}
{Carrillo} A.,  {Deason} A.~J.,  {Fattahi} A.,  {Callingham} T.~M.,   {Grand}
  R. J.~J.,  2023, \mn@doi [arXiv e-prints] {10.48550/arXiv.2306.00770}, \href
  {https://ui.adsabs.harvard.edu/abs/2023arXiv230600770C} {p. arXiv:2306.00770}

\bibitem[\protect\citeauthoryear{{Chabrier}}{{Chabrier}}{2003}]{chabrier_IMF}
{Chabrier} G.,  2003, \mn@doi [\pasp] {10.1086/376392}, \href
  {https://ui.adsabs.harvard.edu/abs/2003PASP..115..763C} {115, 763}

\bibitem[\protect\citeauthoryear{{Chen}, {Ting}  \& {Hayden}}{{Chen}
  et~al.}{2023}]{chen_23}
{Chen} B.,  {Ting} Y.-S.,   {Hayden} M.,  2023, \mn@doi [arXiv e-prints]
  {10.48550/arXiv.2308.15976}, \href
  {https://ui.adsabs.harvard.edu/abs/2023arXiv230815976C} {p. arXiv:2308.15976}

\bibitem[\protect\citeauthoryear{Cleveland}{Cleveland}{1979}]{lowess-orig}
Cleveland W.~S.,  1979, \mn@doi [Journal of the American Statistical
  Association] {10.1080/01621459.1979.10481038}, 74, 829

\bibitem[\protect\citeauthoryear{{Conroy}, {Graves}  \& {van Dokkum}}{{Conroy}
  et~al.}{2014}]{Conroy_etg_14}
{Conroy} C.,  {Graves} G.~J.,   {van Dokkum} P.~G.,  2014, \mn@doi [\apj]
  {10.1088/0004-637X/780/1/33}, \href
  {https://ui.adsabs.harvard.edu/abs/2014ApJ...780...33C} {780, 33}

\bibitem[\protect\citeauthoryear{{Conroy} et~al.,}{{Conroy} et~al.}{2019}]{H3}
{Conroy} C.,  et~al., 2019, \mn@doi [\apj] {10.3847/1538-4357/ab38b8}, \href
  {https://ui.adsabs.harvard.edu/abs/2019ApJ...883..107C} {883, 107}

\bibitem[\protect\citeauthoryear{{Conroy} et~al.,}{{Conroy}
  et~al.}{2022}]{conroy_h3_23}
{Conroy} C.,  et~al., 2022, \mn@doi [arXiv e-prints]
  {10.48550/arXiv.2204.02989}, \href
  {https://ui.adsabs.harvard.edu/abs/2022arXiv220402989C} {p. arXiv:2204.02989}

\bibitem[\protect\citeauthoryear{{Cowie}, {Songaila}, {Hu}  \& {Cohen}}{{Cowie}
  et~al.}{1996}]{cowie-downsizing}
{Cowie} L.~L.,  {Songaila} A.,  {Hu} E.~M.,   {Cohen} J.~G.,  1996, \mn@doi
  [\aj] {10.1086/118058}, \href
  {https://ui.adsabs.harvard.edu/abs/1996AJ....112..839C} {112, 839}

\bibitem[\protect\citeauthoryear{{Crain} \& {van de Voort}}{{Crain} \& {van de
  Voort}}{2023}]{Crain_and_van_de_Voort_23}
{Crain} R.~A.,  {van de Voort} F.,  2023, \mn@doi [\araa]
  {10.1146/annurev-astro-041923-043618}, \href
  {https://ui.adsabs.harvard.edu/abs/2023ARA&A..61..473C} {61, 473}

\bibitem[\protect\citeauthoryear{{Crain} et~al.,}{{Crain}
  et~al.}{2015}]{crain-calibration}
{Crain} R.~A.,  et~al., 2015, \mn@doi [\mnras] {10.1093/mnras/stv725}, \href
  {https://ui.adsabs.harvard.edu/abs/2015MNRAS.450.1937C} {450, 1937}

\bibitem[\protect\citeauthoryear{{Cullen} \& {Dehnen}}{{Cullen} \&
  {Dehnen}}{2010}]{cullen-inviscid}
{Cullen} L.,  {Dehnen} W.,  2010, \mn@doi [\mnras]
  {10.1111/j.1365-2966.2010.17158.x}, \href
  {https://ui.adsabs.harvard.edu/abs/2010MNRAS.408..669C} {408, 669}

\bibitem[\protect\citeauthoryear{{Dalla Vecchia} \& {Schaye}}{{Dalla Vecchia}
  \& {Schaye}}{2012}]{schaye-thermal-feedback}
{Dalla Vecchia} C.,  {Schaye} J.,  2012, \mn@doi [\mnras]
  {10.1111/j.1365-2966.2012.21704.x}, \href
  {https://ui.adsabs.harvard.edu/abs/2012MNRAS.426..140D} {426, 140}

\bibitem[\protect\citeauthoryear{{Davies}, {Crain}, {McCarthy}, {Oppenheimer},
  {Schaye}, {Schaller}  \& {McAlpine}}{{Davies} et~al.}{2019}]{davies-lowess_1}
{Davies} J.~J.,  {Crain} R.~A.,  {McCarthy} I.~G.,  {Oppenheimer} B.~D.,
  {Schaye} J.,  {Schaller} M.,   {McAlpine} S.,  2019, \mn@doi [\mnras]
  {10.1093/mnras/stz635}, \href
  {https://ui.adsabs.harvard.edu/abs/2019MNRAS.485.3783D} {485, 3783}

\bibitem[\protect\citeauthoryear{{Davies}, {Crain}, {Oppenheimer}  \&
  {Schaye}}{{Davies} et~al.}{2020}]{davies_lowess_2}
{Davies} J.~J.,  {Crain} R.~A.,  {Oppenheimer} B.~D.,   {Schaye} J.,  2020,
  \mn@doi [\mnras] {10.1093/mnras/stz3201}, \href
  {https://ui.adsabs.harvard.edu/abs/2020MNRAS.491.4462D} {491, 4462}

\bibitem[\protect\citeauthoryear{{De Silva} et~al.,}{{De Silva}
  et~al.}{2015}]{GALAH}
{De Silva} G.~M.,  et~al., 2015, \mn@doi [\mnras] {10.1093/mnras/stv327}, \href
  {https://ui.adsabs.harvard.edu/abs/2015MNRAS.449.2604D} {449, 2604}

\bibitem[\protect\citeauthoryear{{Dolag}, {Borgani}, {Schindler}, {Diaferio}
  \& {Bykov}}{{Dolag} et~al.}{2008}]{dolag-subfind}
{Dolag} K.,  {Borgani} S.,  {Schindler} S.,  {Diaferio} A.,   {Bykov} A.~M.,
  2008, \mn@doi [\ssr] {10.1007/s11214-008-9316-5}, \href
  {https://ui.adsabs.harvard.edu/abs/2008SSRv..134..229D} {134, 229}

\bibitem[\protect\citeauthoryear{{Drlica-Wagner} et~al.,}{{Drlica-Wagner}
  et~al.}{2015}]{drlica-lgds}
{Drlica-Wagner} A.,  et~al., 2015, \mn@doi [\apj]
  {10.1088/0004-637X/813/2/109}, \href
  {https://ui.adsabs.harvard.edu/abs/2015ApJ...813..109D} {813, 109}

\bibitem[\protect\citeauthoryear{{Durier} \& {Dalla Vecchia}}{{Durier} \&
  {Dalla Vecchia}}{2012}]{durier-timestep}
{Durier} F.,  {Dalla Vecchia} C.,  2012, \mn@doi [\mnras]
  {10.1111/j.1365-2966.2011.19712.x}, \href
  {https://ui.adsabs.harvard.edu/abs/2012MNRAS.419..465D} {419, 465}

\bibitem[\protect\citeauthoryear{{Fernandes} et~al.,}{{Fernandes}
  et~al.}{2023}]{Fernandes2023}
{Fernandes} L.,  et~al., 2023, \mn@doi [\mnras] {10.1093/mnras/stac3543}, \href
  {https://ui.adsabs.harvard.edu/abs/2023MNRAS.519.3611F} {519, 3611}

\bibitem[\protect\citeauthoryear{{Gaia Collaboration} et~al.,}{{Gaia
  Collaboration} et~al.}{2018}]{gaia}
{Gaia Collaboration} et~al., 2018, \mn@doi [\aap]
  {10.1051/0004-6361/201833051}, \href
  {https://ui.adsabs.harvard.edu/abs/2018A&A...616A...1G} {616, A1}

\bibitem[\protect\citeauthoryear{{Gallart} et~al.,}{{Gallart}
  et~al.}{2015}]{dwarfs-sfh-3-gallart}
{Gallart} C.,  et~al., 2015, \mn@doi [\apjl] {10.1088/2041-8205/811/2/L18},
  \href {https://ui.adsabs.harvard.edu/abs/2015ApJ...811L..18G} {811, L18}

\bibitem[\protect\citeauthoryear{{Gebek} \& {Matthee}}{{Gebek} \&
  {Matthee}}{2022}]{gebek-alpha-eagle}
{Gebek} A.,  {Matthee} J.,  2022, \mn@doi [\apj] {10.3847/1538-4357/ac350b},
  \href {https://ui.adsabs.harvard.edu/abs/2022ApJ...924...73G} {924, 73}

\bibitem[\protect\citeauthoryear{{Gilmore} \& {Wyse}}{{Gilmore} \&
  {Wyse}}{1991}]{gilmore_91_sne}
{Gilmore} G.,  {Wyse} R. F.~G.,  1991, \mn@doi [\apjl] {10.1086/185930}, \href
  {https://ui.adsabs.harvard.edu/abs/1991ApJ...367L..55G} {367, L55}

\bibitem[\protect\citeauthoryear{{Gilmore} et~al.,}{{Gilmore}
  et~al.}{2012}]{gilmore-geso}
{Gilmore} G.,  et~al., 2012, The Messenger, \href
  {https://ui.adsabs.harvard.edu/abs/2012Msngr.147...25G} {147, 25}

\bibitem[\protect\citeauthoryear{{Gonzalez} et~al.,}{{Gonzalez}
  et~al.}{2011}]{gonzales-11-afe}
{Gonzalez} O.~A.,  et~al., 2011, \mn@doi [\aap] {10.1051/0004-6361/201116548},
  \href {https://ui.adsabs.harvard.edu/abs/2011A&A...530A..54G} {530, A54}

\bibitem[\protect\citeauthoryear{{Gonzalez} et~al.,}{{Gonzalez}
  et~al.}{2020}]{MOONS2}
{Gonzalez} O.~A.,  et~al., 2020, \mn@doi [The Messenger]
  {10.18727/0722-6691/5196}, \href
  {https://ui.adsabs.harvard.edu/abs/2020Msngr.180...18G} {180, 18}

\bibitem[\protect\citeauthoryear{{Graur} \& {Maoz}}{{Graur} \&
  {Maoz}}{2013}]{graur-dtd}
{Graur} O.,  {Maoz} D.,  2013, \mn@doi [\mnras] {10.1093/mnras/sts718}, \href
  {https://ui.adsabs.harvard.edu/abs/2013MNRAS.430.1746G} {430, 1746}

\bibitem[\protect\citeauthoryear{{Griffith}, {Sukhbold}, {Weinberg}, {Johnson},
  {Johnson}  \& {Vincenzo}}{{Griffith} et~al.}{2021}]{vice3}
{Griffith} E.~J.,  {Sukhbold} T.,  {Weinberg} D.~H.,  {Johnson} J.~A.,
  {Johnson} J.~W.,   {Vincenzo} F.,  2021, arXiv e-prints, \href
  {https://ui.adsabs.harvard.edu/abs/2021arXiv210309837G} {p. arXiv:2103.09837}

\bibitem[\protect\citeauthoryear{{Gronow}, {Collins}, {Sim}  \&
  {R{\"o}pke}}{{Gronow} et~al.}{2021a}]{gronow21a}
{Gronow} S.,  {Collins} C.~E.,  {Sim} S.~A.,   {R{\"o}pke} F.~K.,  2021a,
  \mn@doi [\aap] {10.1051/0004-6361/202039954}, \href
  {https://ui.adsabs.harvard.edu/abs/2021A&A...649A.155G} {649, A155}

\bibitem[\protect\citeauthoryear{{Gronow}, {C{\^o}t{\'e}}, {Lach},
  {Seitenzahl}, {Collins}, {Sim}  \& {R{\"o}pke}}{{Gronow}
  et~al.}{2021b}]{gronow21b}
{Gronow} S.,  {C{\^o}t{\'e}} B.,  {Lach} F.,  {Seitenzahl} I.~R.,  {Collins}
  C.~E.,  {Sim} S.~A.,   {R{\"o}pke} F.~K.,  2021b, \mn@doi [\aap]
  {10.1051/0004-6361/202140881}, \href
  {https://ui.adsabs.harvard.edu/abs/2021A&A...656A..94G} {656, A94}

\bibitem[\protect\citeauthoryear{{Hasselquist} et~al.,}{{Hasselquist}
  et~al.}{2021}]{hasselquist-dwarfs}
{Hasselquist} S.,  et~al., 2021, \mn@doi [\apj] {10.3847/1538-4357/ac25f9},
  \href {https://ui.adsabs.harvard.edu/abs/2021ApJ...923..172H} {923, 172}

\bibitem[\protect\citeauthoryear{{Hayden} et~al.,}{{Hayden}
  et~al.}{2015}]{hayden-2015}
{Hayden} M.~R.,  et~al., 2015, \mn@doi [\apj] {10.1088/0004-637X/808/2/132},
  \href {https://ui.adsabs.harvard.edu/abs/2015ApJ...808..132H} {808, 132}

\bibitem[\protect\citeauthoryear{{Hendricks}, {Koch}, {Lanfranchi}, {Boeche},
  {Walker}, {Johnson}, {Pe{\~n}arrubia}  \& {Gilmore}}{{Hendricks}
  et~al.}{2014}]{hendricks-knee}
{Hendricks} B.,  {Koch} A.,  {Lanfranchi} G.~A.,  {Boeche} C.,  {Walker} M.,
  {Johnson} C.~I.,  {Pe{\~n}arrubia} J.,   {Gilmore} G.,  2014, \mn@doi [\apj]
  {10.1088/0004-637X/785/2/102}, \href
  {https://ui.adsabs.harvard.edu/abs/2014ApJ...785..102H} {785, 102}

\bibitem[\protect\citeauthoryear{Hoffman, Gelman  et~al.}{Hoffman
  et~al.}{2014}]{NUTS-sampler}
Hoffman M.~D.,  Gelman A.,   et~al., 2014, J. Mach. Learn. Res., 15, 1593

\bibitem[\protect\citeauthoryear{{Hogg}, {Bovy}  \& {Lang}}{{Hogg}
  et~al.}{2010}]{hogg-pw}
{Hogg} D.~W.,  {Bovy} J.,   {Lang} D.,  2010, arXiv e-prints, \href
  {https://ui.adsabs.harvard.edu/abs/2010arXiv1008.4686H} {p. arXiv:1008.4686}

\bibitem[\protect\citeauthoryear{{Hopkins}}{{Hopkins}}{2013}]{hopkins-sph}
{Hopkins} P.~F.,  2013, \mn@doi [\mnras] {10.1093/mnras/sts210}, \href
  {https://ui.adsabs.harvard.edu/abs/2013MNRAS.428.2840H} {428, 2840}

\bibitem[\protect\citeauthoryear{{Horta} et~al.,}{{Horta}
  et~al.}{2021}]{horta-heracles}
{Horta} D.,  et~al., 2021, \mn@doi [\mnras] {10.1093/mnras/staa2987}, \href
  {https://ui.adsabs.harvard.edu/abs/2021MNRAS.500.1385H} {500, 1385}

\bibitem[\protect\citeauthoryear{{Horta} et~al.,}{{Horta}
  et~al.}{2023}]{horta-substructure}
{Horta} D.,  et~al., 2023, \mn@doi [\mnras] {10.1093/mnras/stac3179}, \href
  {https://ui.adsabs.harvard.edu/abs/2023MNRAS.520.5671H} {520, 5671}

\bibitem[\protect\citeauthoryear{{Hughes}, {Pfeffer}, {Martig}, {Reina-Campos},
  {Bastian}, {Crain}  \& {Kruijssen}}{{Hughes} et~al.}{2020}]{hughes-gc}
{Hughes} M.~E.,  {Pfeffer} J.~L.,  {Martig} M.,  {Reina-Campos} M.,  {Bastian}
  N.,  {Crain} R.~A.,   {Kruijssen} J.~M.~D.,  2020, \mn@doi [\mnras]
  {10.1093/mnras/stz3341}, \href
  {https://ui.adsabs.harvard.edu/abs/2020MNRAS.491.4012H} {491, 4012}

\bibitem[\protect\citeauthoryear{Hunter}{Hunter}{2007}]{hunter_matplotlib}
Hunter J.~D.,  2007, \mn@doi [Computing in Science & Engineering]
  {10.1109/MCSE.2007.55}, 9, 90

\bibitem[\protect\citeauthoryear{{Iwamoto}, {Brachwitz}, {Nomoto}, {Kishimoto},
  {Umeda}, {Hix}  \& {Thielemann}}{{Iwamoto} et~al.}{1999}]{iwamoto_1a}
{Iwamoto} K.,  {Brachwitz} F.,  {Nomoto} K.,  {Kishimoto} N.,  {Umeda} H.,
  {Hix} W.~R.,   {Thielemann} F.-K.,  1999, \mn@doi [\apjs] {10.1086/313278},
  \href {https://ui.adsabs.harvard.edu/abs/1999ApJS..125..439I} {125, 439}

\bibitem[\protect\citeauthoryear{{Jin} et~al.,}{{Jin} et~al.}{2023}]{weave}
{Jin} S.,  et~al., 2023, \mn@doi [\mnras] {10.1093/mnras/stad557}, \href
  {https://ui.adsabs.harvard.edu/abs/2023MNRAS.tmp..715J} {}

\bibitem[\protect\citeauthoryear{{Johansson} et~al.,}{{Johansson}
  et~al.}{2013}]{Johansson12}
{Johansson} J.,  et~al., 2013, \mn@doi [\mnras] {10.1093/mnras/stt1408}, \href
  {https://ui.adsabs.harvard.edu/abs/2013MNRAS.435.1680J} {435, 1680}

\bibitem[\protect\citeauthoryear{{Johnson} \& {Weinberg}}{{Johnson} \&
  {Weinberg}}{2020}]{vice1}
{Johnson} J.~W.,  {Weinberg} D.~H.,  2020, \mn@doi [\mnras]
  {10.1093/mnras/staa2431}, \href
  {https://ui.adsabs.harvard.edu/abs/2020MNRAS.498.1364J} {498, 1364}

\bibitem[\protect\citeauthoryear{{Johnson} et~al.,}{{Johnson}
  et~al.}{2021}]{vice2}
{Johnson} J.~W.,  et~al., 2021, arXiv e-prints, \href
  {https://ui.adsabs.harvard.edu/abs/2021arXiv210309838J} {p. arXiv:2103.09838}

\bibitem[\protect\citeauthoryear{{Keller}, {Wadsley}, {Benincasa}  \&
  {Couchman}}{{Keller} et~al.}{2014}]{keller-14}
{Keller} B.~W.,  {Wadsley} J.,  {Benincasa} S.~M.,   {Couchman} H.~M.~P.,
  2014, \mn@doi [\mnras] {10.1093/mnras/stu1058}, \href
  {https://ui.adsabs.harvard.edu/abs/2014MNRAS.442.3013K} {442, 3013}

\bibitem[\protect\citeauthoryear{{Kennicutt}}{{Kennicutt}}{1998}]{KS-98}
{Kennicutt} Robert~C. J.,  1998, \mn@doi [\apj] {10.1086/305588}, \href
  {https://ui.adsabs.harvard.edu/abs/1998ApJ...498..541K} {498, 541}

\bibitem[\protect\citeauthoryear{{Kirby}, {Cohen}, {Guhathakurta}, {Cheng},
  {Bullock}  \& {Gallazzi}}{{Kirby} et~al.}{2013}]{kirby13}
{Kirby} E.~N.,  {Cohen} J.~G.,  {Guhathakurta} P.,  {Cheng} L.,  {Bullock}
  J.~S.,   {Gallazzi} A.,  2013, \mn@doi [\apj] {10.1088/0004-637X/779/2/102},
  \href {https://ui.adsabs.harvard.edu/abs/2013ApJ...779..102K} {779, 102}

\bibitem[\protect\citeauthoryear{{Kobayashi}, {Umeda}, {Nomoto}, {Tominaga}  \&
  {Ohkubo}}{{Kobayashi} et~al.}{2006}]{kobayashi_06}
{Kobayashi} C.,  {Umeda} H.,  {Nomoto} K.,  {Tominaga} N.,   {Ohkubo} T.,
  2006, \mn@doi [\apj] {10.1086/508914}, \href
  {https://ui.adsabs.harvard.edu/abs/2006ApJ...653.1145K} {653, 1145}

\bibitem[\protect\citeauthoryear{{Kroupa}}{{Kroupa}}{2001}]{kroupa}
{Kroupa} P.,  2001, \mn@doi [\mnras] {10.1046/j.1365-8711.2001.04022.x}, \href
  {https://ui.adsabs.harvard.edu/abs/2001MNRAS.322..231K} {322, 231}

\bibitem[\protect\citeauthoryear{{Kruijssen}, {Pfeffer}, {Crain}  \&
  {Bastian}}{{Kruijssen} et~al.}{2019}]{kruijssen-emosaics}
{Kruijssen} J.~M.~D.,  {Pfeffer} J.~L.,  {Crain} R.~A.,   {Bastian} N.,  2019,
  \mn@doi [\mnras] {10.1093/mnras/stz968}, \href
  {https://ui.adsabs.harvard.edu/abs/2019MNRAS.486.3134K} {486, 3134}

\bibitem[\protect\citeauthoryear{{Leroy}, {Walter}, {Brinks}, {Bigiel}, {de
  Blok}, {Madore}  \& {Thornley}}{{Leroy} et~al.}{2008}]{leroy08}
{Leroy} A.~K.,  {Walter} F.,  {Brinks} E.,  {Bigiel} F.,  {de Blok} W.~J.~G.,
  {Madore} B.,   {Thornley} M.~D.,  2008, \mn@doi [\aj]
  {10.1088/0004-6256/136/6/2782}, \href
  {https://ui.adsabs.harvard.edu/abs/2008AJ....136.2782L} {136, 2782}

\bibitem[\protect\citeauthoryear{{Leroy} et~al.,}{{Leroy}
  et~al.}{2013}]{leroy_13}
{Leroy} A.~K.,  et~al., 2013, \mn@doi [\aj] {10.1088/0004-6256/146/2/19}, \href
  {https://ui.adsabs.harvard.edu/abs/2013AJ....146...19L} {146, 19}

\bibitem[\protect\citeauthoryear{{Li} \& {White}}{{Li} \&
  {White}}{2009}]{li-gsmf}
{Li} C.,  {White} S. D.~M.,  2009, \mn@doi [\mnras]
  {10.1111/j.1365-2966.2009.15268.x}, \href
  {https://ui.adsabs.harvard.edu/abs/2009MNRAS.398.2177L} {398, 2177}

\bibitem[\protect\citeauthoryear{{Mackereth}, {Crain}, {Schiavon}, {Schaye},
  {Theuns}  \& {Schaller}}{{Mackereth} et~al.}{2018}]{mackereth-bimodality}
{Mackereth} J.~T.,  {Crain} R.~A.,  {Schiavon} R.~P.,  {Schaye} J.,  {Theuns}
  T.,   {Schaller} M.,  2018, \mn@doi [\mnras] {10.1093/mnras/sty972}, \href
  {https://ui.adsabs.harvard.edu/abs/2018MNRAS.477.5072M} {477, 5072}

\bibitem[\protect\citeauthoryear{{Mackereth} et~al.,}{{Mackereth}
  et~al.}{2019}]{mackereth-accreted}
{Mackereth} J.~T.,  et~al., 2019, \mn@doi [\mnras] {10.1093/mnras/sty2955},
  \href {https://ui.adsabs.harvard.edu/abs/2019MNRAS.482.3426M} {482, 3426}

\bibitem[\protect\citeauthoryear{{Madau} \& {Fragos}}{{Madau} \&
  {Fragos}}{2017}]{madau_17_dtd}
{Madau} P.,  {Fragos} T.,  2017, \mn@doi [\apj] {10.3847/1538-4357/aa6af9},
  \href {https://ui.adsabs.harvard.edu/abs/2017ApJ...840...39M} {840, 39}

\bibitem[\protect\citeauthoryear{{Majewski} et~al.,}{{Majewski}
  et~al.}{2017}]{apogee}
{Majewski} S.~R.,  et~al., 2017, \mn@doi [\aj] {10.3847/1538-3881/aa784d},
  \href {https://ui.adsabs.harvard.edu/abs/2017AJ....154...94M} {154, 94}

\bibitem[\protect\citeauthoryear{{Maoz} \& {Graur}}{{Maoz} \&
  {Graur}}{2017}]{Maoz-Graur-Alpha}
{Maoz} D.,  {Graur} O.,  2017, \mn@doi [\apj] {10.3847/1538-4357/aa8b6e}, \href
  {https://ui.adsabs.harvard.edu/abs/2017ApJ...848...25M} {848, 25}

\bibitem[\protect\citeauthoryear{{Marigo}}{{Marigo}}{2001}]{marigo-yields}
{Marigo} P.,  2001, \mn@doi [\aap] {10.1051/0004-6361:20000247}, \href
  {https://ui.adsabs.harvard.edu/abs/2001A&A...370..194M} {370, 194}

\bibitem[\protect\citeauthoryear{{Martell} et~al.,}{{Martell}
  et~al.}{2017}]{GALAH2}
{Martell} S.~L.,  et~al., 2017, \mn@doi [\mnras] {10.1093/mnras/stw2835}, \href
  {https://ui.adsabs.harvard.edu/abs/2017MNRAS.465.3203M} {465, 3203}

\bibitem[\protect\citeauthoryear{{Matteucci} \& {Brocato}}{{Matteucci} \&
  {Brocato}}{1990}]{matteucci-chemistry}
{Matteucci} F.,  {Brocato} E.,  1990, \mn@doi [\apj] {10.1086/169508}, \href
  {https://ui.adsabs.harvard.edu/abs/1990ApJ...365..539M} {365, 539}

\bibitem[\protect\citeauthoryear{{Matteucci}, {Spitoni}, {Recchi}  \&
  {Valiante}}{{Matteucci} et~al.}{2009}]{matteucci-ias}
{Matteucci} F.,  {Spitoni} E.,  {Recchi} S.,   {Valiante} R.,  2009, \mn@doi
  [\aap] {10.1051/0004-6361/200911869}, \href
  {https://ui.adsabs.harvard.edu/abs/2009A&A...501..531M} {501, 531}

\bibitem[\protect\citeauthoryear{{McAlpine} et~al.,}{{McAlpine}
  et~al.}{2016}]{mcalpine-data-release}
{McAlpine} S.,  et~al., 2016, \mn@doi [Astronomy and Computing]
  {10.1016/j.ascom.2016.02.004}, \href
  {https://ui.adsabs.harvard.edu/abs/2016A&C....15...72M} {15, 72}

\bibitem[\protect\citeauthoryear{{McConnachie}}{{McConnachie}}{2012}]{mcconachie-12}
{McConnachie} A.~W.,  2012, \mn@doi [\aj] {10.1088/0004-6256/144/1/4}, \href
  {https://ui.adsabs.harvard.edu/abs/2012AJ....144....4M} {144, 4}

\bibitem[\protect\citeauthoryear{{McWilliam}}{{McWilliam}}{1997}]{McWilliam_MW}
{McWilliam} A.,  1997, \mn@doi [\araa] {10.1146/annurev.astro.35.1.503}, \href
  {https://ui.adsabs.harvard.edu/abs/1997ARA&A..35..503M} {35, 503}

\bibitem[\protect\citeauthoryear{{Mitchell}, {Schaye}, {Bower}  \&
  {Crain}}{{Mitchell} et~al.}{2020}]{mitchell_20}
{Mitchell} P.~D.,  {Schaye} J.,  {Bower} R.~G.,   {Crain} R.~A.,  2020, \mn@doi
  [\mnras] {10.1093/mnras/staa938}, \href
  {https://ui.adsabs.harvard.edu/abs/2020MNRAS.494.3971M} {494, 3971}

\bibitem[\protect\citeauthoryear{{Newton}, {Cautun}, {Jenkins}, {Frenk}  \&
  {Helly}}{{Newton} et~al.}{2018}]{newton-lgds}
{Newton} O.,  {Cautun} M.,  {Jenkins} A.,  {Frenk} C.~S.,   {Helly} J.~C.,
  2018, \mn@doi [\mnras] {10.1093/mnras/sty1085}, \href
  {https://ui.adsabs.harvard.edu/abs/2018MNRAS.479.2853N} {479, 2853}

\bibitem[\protect\citeauthoryear{{Nidever} et~al.,}{{Nidever}
  et~al.}{2020}]{nidever-lazy-giants}
{Nidever} D.~L.,  et~al., 2020, \mn@doi [\apj] {10.3847/1538-4357/ab7305},
  \href {https://ui.adsabs.harvard.edu/abs/2020ApJ...895...88N} {895, 88}

\bibitem[\protect\citeauthoryear{{Nidever} et~al.,}{{Nidever}
  et~al.}{2021}]{nidever-smash}
{Nidever} D.~L.,  et~al., 2021, \mn@doi [\aj] {10.3847/1538-3881/abceb7}, \href
  {https://ui.adsabs.harvard.edu/abs/2021AJ....161...74N} {161, 74}

\bibitem[\protect\citeauthoryear{Perez \& Granger}{Perez \&
  Granger}{2007}]{perez_ipython}
Perez F.,  Granger B.~E.,  2007, \mn@doi [Computing in Science & Engineering]
  {10.1109/MCSE.2007.53}, 9, 21

\bibitem[\protect\citeauthoryear{{Pfeffer}, {Kruijssen}, {Crain}  \&
  {Bastian}}{{Pfeffer} et~al.}{2018}]{pfeffer-emosaics}
{Pfeffer} J.,  {Kruijssen} J.~M.~D.,  {Crain} R.~A.,   {Bastian} N.,  2018,
  \mn@doi [\mnras] {10.1093/mnras/stx3124}, \href
  {https://ui.adsabs.harvard.edu/abs/2018MNRAS.475.4309P} {475, 4309}

\bibitem[\protect\citeauthoryear{{Philcox} \& {Rybizki}}{{Philcox} \&
  {Rybizki}}{2019}]{chempy}
{Philcox} O. H.~E.,  {Rybizki} J.,  2019, \mn@doi [\apj]
  {10.3847/1538-4357/ab5186}, \href
  {https://ui.adsabs.harvard.edu/abs/2019ApJ...887....9P} {887, 9}

\bibitem[\protect\citeauthoryear{{Planck Collaboration} et~al.,}{{Planck
  Collaboration} et~al.}{2014}]{planck_results}
{Planck Collaboration} et~al., 2014, \mn@doi [\aap]
  {10.1051/0004-6361/201321529}, \href
  {https://ui.adsabs.harvard.edu/abs/2014A&A...571A...1P} {571, A1}

\bibitem[\protect\citeauthoryear{{Portinari}, {Chiosi}  \&
  {Bressan}}{{Portinari} et~al.}{1998}]{portinari-yields}
{Portinari} L.,  {Chiosi} C.,   {Bressan} A.,  1998, \aap, \href
  {https://ui.adsabs.harvard.edu/abs/1998A&A...334..505P} {334, 505}

\bibitem[\protect\citeauthoryear{{Price}}{{Price}}{2010}]{price}
{Price} D.~J.,  2010, \mn@doi [\mnras] {10.1111/j.1365-2966.2009.15763.x},
  \href {https://ui.adsabs.harvard.edu/abs/2010MNRAS.401.1475P} {401, 1475}

\bibitem[\protect\citeauthoryear{{Qu} et~al.,}{{Qu} et~al.}{2017}]{qu_gama_17}
{Qu} Y.,  et~al., 2017, \mn@doi [\mnras] {10.1093/mnras/stw2437}, \href
  {https://ui.adsabs.harvard.edu/abs/2017MNRAS.464.1659Q} {464, 1659}

\bibitem[\protect\citeauthoryear{{Randich}, {Gilmore}  \& {Gaia-ESO
  Consortium}}{{Randich} et~al.}{2013}]{randich-geso}
{Randich} S.,  {Gilmore} G.,   {Gaia-ESO Consortium} 2013, The Messenger, \href
  {https://ui.adsabs.harvard.edu/abs/2013Msngr.154...47R} {154, 47}

\bibitem[\protect\citeauthoryear{{Rosas-Guevara} et~al.,}{{Rosas-Guevara}
  et~al.}{2015}]{rosas-2015}
{Rosas-Guevara} Y.~M.,  et~al., 2015, \mn@doi [\mnras] {10.1093/mnras/stv2056},
  \href {https://ui.adsabs.harvard.edu/abs/2015MNRAS.454.1038R} {454, 1038}

\bibitem[\protect\citeauthoryear{Salvatier, Wiecki  \& Fonnesbeck}{Salvatier
  et~al.}{2016}]{pymc3}
Salvatier J.,  Wiecki T.~V.,   Fonnesbeck C.,  2016, PeerJ Computer Science, 2,
  e55

\bibitem[\protect\citeauthoryear{{Schaye}}{{Schaye}}{2004}]{schayethreshold}
{Schaye} J.,  2004, \mn@doi [\apj] {10.1086/421232}, \href
  {https://ui.adsabs.harvard.edu/abs/2004ApJ...609..667S} {609, 667}

\bibitem[\protect\citeauthoryear{{Schaye} \& {Dalla Vecchia}}{{Schaye} \&
  {Dalla Vecchia}}{2008}]{schaye-ks-law}
{Schaye} J.,  {Dalla Vecchia} C.,  2008, \mn@doi [\mnras]
  {10.1111/j.1365-2966.2007.12639.x}, \href
  {https://ui.adsabs.harvard.edu/abs/2008MNRAS.383.1210S} {383, 1210}

\bibitem[\protect\citeauthoryear{{Schaye} et~al.,}{{Schaye}
  et~al.}{2015}]{schaye-eagle}
{Schaye} J.,  et~al., 2015, \mn@doi [\mnras] {10.1093/mnras/stu2058}, \href
  {https://ui.adsabs.harvard.edu/abs/2015MNRAS.446..521S} {446, 521}

\bibitem[\protect\citeauthoryear{{Schaye} et~al.,}{{Schaye}
  et~al.}{2023}]{schaye_flamingo}
{Schaye} J.,  et~al., 2023, \mn@doi [\mnras] {10.1093/mnras/stad2419}, \href
  {https://ui.adsabs.harvard.edu/abs/2023MNRAS.526.4978S} {526, 4978}

\bibitem[\protect\citeauthoryear{{Schiavon}}{{Schiavon}}{2007}]{Schiavon2007}
{Schiavon} R.~P.,  2007, \mn@doi [\apjs] {10.1086/511753}, \href
  {https://ui.adsabs.harvard.edu/abs/2007ApJS..171..146S} {171, 146}

\bibitem[\protect\citeauthoryear{{Schiavon}, {Mackereth}, {Pfeffer}, {Crain}
  \& {Bovy}}{{Schiavon} et~al.}{2020}]{ripisc-MW-symposium}
{Schiavon} R.~P.,  {Mackereth} J.~T.,  {Pfeffer} J.,  {Crain} R.~A.,   {Bovy}
  J.,  2020, in {Bragaglia} A.,  {Davies} M.,  {Sills} A.,   {Vesperini} E.,
  eds, Star Clusters: From the Milky Way to the Early Universe. 351.
pp 170--173 (\mn@eprint {arXiv} {2002.08380}),
  \mn@doi{10.1017/S1743921319007889}

\bibitem[\protect\citeauthoryear{Seabold \& Perktold}{Seabold \&
  Perktold}{2010}]{seabold_statsmodels}
Seabold S.,  Perktold J.,  2010, in 9th Python in Science Conference.

\bibitem[\protect\citeauthoryear{{Segers}, {Schaye}, {Bower}, {Crain},
  {Schaller}  \& {Theuns}}{{Segers} et~al.}{2016}]{segers-alpha}
{Segers} M.~C.,  {Schaye} J.,  {Bower} R.~G.,  {Crain} R.~A.,  {Schaller} M.,
  {Theuns} T.,  2016, \mn@doi [\mnras] {10.1093/mnrasl/slw111}, \href
  {https://ui.adsabs.harvard.edu/abs/2016MNRAS.461L.102S} {461, L102}

\bibitem[\protect\citeauthoryear{{Seitenzahl} et~al.,}{{Seitenzahl}
  et~al.}{2013}]{seitenzahl13}
{Seitenzahl} I.~R.,  et~al., 2013, \mn@doi [\mnras] {10.1093/mnras/sts402},
  \href {https://ui.adsabs.harvard.edu/abs/2013MNRAS.429.1156S} {429, 1156}

\bibitem[\protect\citeauthoryear{{Skillman} et~al.,}{{Skillman}
  et~al.}{2017}]{dwarfs-sfh-4-skillman}
{Skillman} E.~D.,  et~al., 2017, \mn@doi [\apj] {10.3847/1538-4357/aa60c5},
  \href {https://ui.adsabs.harvard.edu/abs/2017ApJ...837..102S} {837, 102}

\bibitem[\protect\citeauthoryear{{Springel}}{{Springel}}{2005}]{springel-gadget}
{Springel} V.,  2005, \mn@doi [\mnras] {10.1111/j.1365-2966.2005.09655.x},
  \href {https://ui.adsabs.harvard.edu/abs/2005MNRAS.364.1105S} {364, 1105}

\bibitem[\protect\citeauthoryear{{Springel}, {White}, {Tormen}  \&
  {Kauffmann}}{{Springel} et~al.}{2001}]{springel-subfind}
{Springel} V.,  {White} S. D.~M.,  {Tormen} G.,   {Kauffmann} G.,  2001,
  \mn@doi [\mnras] {10.1046/j.1365-8711.2001.04912.x}, \href
  {https://ui.adsabs.harvard.edu/abs/2001MNRAS.328..726S} {328, 726}

\bibitem[\protect\citeauthoryear{{Steinmetz} et~al.,}{{Steinmetz}
  et~al.}{2006}]{RAVE}
{Steinmetz} M.,  et~al., 2006, \mn@doi [\aj] {10.1086/506564}, \href
  {https://ui.adsabs.harvard.edu/abs/2006AJ....132.1645S} {132, 1645}

\bibitem[\protect\citeauthoryear{{Sun} et~al.,}{{Sun} et~al.}{2023}]{sun_23}
{Sun} F.,  et~al., 2023, \mn@doi [\apj] {10.3847/1538-4357/acd53c}, \href
  {https://ui.adsabs.harvard.edu/abs/2023ApJ...953...53S} {953, 53}

\bibitem[\protect\citeauthoryear{{Tacconi} et~al.,}{{Tacconi}
  et~al.}{2018}]{tacconi_18}
{Tacconi} L.~J.,  et~al., 2018, \mn@doi [\apj] {10.3847/1538-4357/aaa4b4},
  \href {https://ui.adsabs.harvard.edu/abs/2018ApJ...853..179T} {853, 179}

\bibitem[\protect\citeauthoryear{{Thielemann} et~al.,}{{Thielemann}
  et~al.}{2003}]{thielemanns_03}
{Thielemann} F.~K.,  et~al., 2003, in {Hillebrandt} W.,  {Leibundgut} B.,  eds,
  From Twilight to Highlight: The Physics of Supernovae. p.~331,
  \mn@doi{10.1007/10828549_46}

\bibitem[\protect\citeauthoryear{{Thomas}, {Greggio}  \& {Bender}}{{Thomas}
  et~al.}{1998}]{thomas-imf-yields}
{Thomas} D.,  {Greggio} L.,   {Bender} R.,  1998, \mn@doi [\mnras]
  {10.1046/j.1365-8711.1998.01289.x}, \href
  {https://ui.adsabs.harvard.edu/abs/1998MNRAS.296..119T} {296, 119}

\bibitem[\protect\citeauthoryear{{Tinsley}}{{Tinsley}}{1979}]{tinsley-chemistry}
{Tinsley} B.~M.,  1979, \mn@doi [\apj] {10.1086/157039}, \href
  {https://ui.adsabs.harvard.edu/abs/1979ApJ...229.1046T} {229, 1046}

\bibitem[\protect\citeauthoryear{{Tissera}, {Rosas-Guevara}, {Bower}, {Crain},
  {del P Lagos}, {Schaller}, {Schaye}  \& {Theuns}}{{Tissera}
  et~al.}{2019}]{tissera-oxy}
{Tissera} P.~B.,  {Rosas-Guevara} Y.,  {Bower} R.~G.,  {Crain} R.~A.,  {del P
  Lagos} C.,  {Schaller} M.,  {Schaye} J.,   {Theuns} T.,  2019, \mn@doi
  [\mnras] {10.1093/mnras/sty2817}, \href
  {https://ui.adsabs.harvard.edu/abs/2019MNRAS.482.2208T} {482, 2208}

\bibitem[\protect\citeauthoryear{{Tolstoy}, {Hill}  \& {Tosi}}{{Tolstoy}
  et~al.}{2009}]{tolstoy-09}
{Tolstoy} E.,  {Hill} V.,   {Tosi} M.,  2009, \mn@doi [\araa]
  {10.1146/annurev-astro-082708-101650}, \href
  {https://ui.adsabs.harvard.edu/abs/2009ARA&A..47..371T} {47, 371}

\bibitem[\protect\citeauthoryear{{Trager}, {Faber}, {Worthey}  \&
  {Gonz{\'a}lez}}{{Trager} et~al.}{2000}]{Trager2000}
{Trager} S.~C.,  {Faber} S.~M.,  {Worthey} G.,   {Gonz{\'a}lez} J.~J.,  2000,
  \mn@doi [\aj] {10.1086/301442}, \href
  {https://ui.adsabs.harvard.edu/abs/2000AJ....120..165T} {120, 165}

\bibitem[\protect\citeauthoryear{{Vargas}, {Geha}, {Kirby}  \&
  {Simon}}{{Vargas} et~al.}{2013}]{vargas1-mw}
{Vargas} L.~C.,  {Geha} M.,  {Kirby} E.~N.,   {Simon} J.~D.,  2013, \mn@doi
  [\apj] {10.1088/0004-637X/767/2/134}, \href
  {https://ui.adsabs.harvard.edu/abs/2013ApJ...767..134V} {767, 134}

\bibitem[\protect\citeauthoryear{{Vargas}, {Geha}  \& {Tollerud}}{{Vargas}
  et~al.}{2014}]{vargas2-and}
{Vargas} L.~C.,  {Geha} M.~C.,   {Tollerud} E.~J.,  2014, \mn@doi [\apj]
  {10.1088/0004-637X/790/1/73}, \href
  {https://ui.adsabs.harvard.edu/abs/2014ApJ...790...73V} {790, 73}

\bibitem[\protect\citeauthoryear{{Vasini}, {Matteucci}, {Spitoni}  \&
  {Siegert}}{{Vasini} et~al.}{2023}]{vasini_lmc_23}
{Vasini} A.,  {Matteucci} F.,  {Spitoni} E.,   {Siegert} T.,  2023, \mn@doi
  [\mnras] {10.1093/mnras/stad1440}, \href
  {https://ui.adsabs.harvard.edu/abs/2023MNRAS.523.1153V} {523, 1153}

\bibitem[\protect\citeauthoryear{Virtanen et~al.,}{Virtanen
  et~al.}{2020}]{jones_scipy}
Virtanen P.,  et~al., 2020, \mn@doi [Nature Methods]
  {10.1038/s41592-019-0686-2}, \href {https://rdcu.be/b08Wh} {17, 261}

\bibitem[\protect\citeauthoryear{{Wallerstein}}{{Wallerstein}}{1962}]{wallerstein_62}
{Wallerstein} G.,  1962, \mn@doi [\apjs] {10.1086/190067}, \href
  {https://ui.adsabs.harvard.edu/abs/1962ApJS....6..407W} {6, 407}

\bibitem[\protect\citeauthoryear{{Weinberg}, {Andrews}  \&
  {Freudenburg}}{{Weinberg} et~al.}{2017}]{weinberg-sudden-events}
{Weinberg} D.~H.,  {Andrews} B.~H.,   {Freudenburg} J.,  2017, \mn@doi [\apj]
  {10.3847/1538-4357/837/2/183}, \href
  {https://ui.adsabs.harvard.edu/abs/2017ApJ...837..183W} {837, 183}

\bibitem[\protect\citeauthoryear{{Weisz} et~al.,}{{Weisz}
  et~al.}{2011}]{dwarfs-sfh-1-weisz}
{Weisz} D.~R.,  et~al., 2011, \mn@doi [\apj] {10.1088/0004-637X/739/1/5}, \href
  {https://ui.adsabs.harvard.edu/abs/2011ApJ...739....5W} {739, 5}

\bibitem[\protect\citeauthoryear{{Weisz}, {Dolphin}, {Skillman}, {Holtzman},
  {Gilbert}, {Dalcanton}  \& {Williams}}{{Weisz}
  et~al.}{2014}]{dwarfs-sfh-2-weisz}
{Weisz} D.~R.,  {Dolphin} A.~E.,  {Skillman} E.~D.,  {Holtzman} J.,  {Gilbert}
  K.~M.,  {Dalcanton} J.~J.,   {Williams} B.~F.,  2014, \mn@doi [\apj]
  {10.1088/0004-637X/789/2/147}, \href
  {https://ui.adsabs.harvard.edu/abs/2014ApJ...789..147W} {789, 147}

\bibitem[\protect\citeauthoryear{{Wiersma}, {Schaye}  \& {Smith}}{{Wiersma}
  et~al.}{2009a}]{wiersma-photocooling}
{Wiersma} R. P.~C.,  {Schaye} J.,   {Smith} B.~D.,  2009a, \mn@doi [\mnras]
  {10.1111/j.1365-2966.2008.14191.x}, \href
  {https://ui.adsabs.harvard.edu/abs/2009MNRAS.393...99W} {393, 99}

\bibitem[\protect\citeauthoryear{{Wiersma}, {Schaye}, {Theuns}, {Dalla Vecchia}
   \& {Tornatore}}{{Wiersma} et~al.}{2009b}]{wiersma-chemistry}
{Wiersma} R. P.~C.,  {Schaye} J.,  {Theuns} T.,  {Dalla Vecchia} C.,
  {Tornatore} L.,  2009b, \mn@doi [\mnras] {10.1111/j.1365-2966.2009.15331.x},
  \href {https://ui.adsabs.harvard.edu/abs/2009MNRAS.399..574W} {399, 574}

\bibitem[\protect\citeauthoryear{{Woosley} \& {Weaver}}{{Woosley} \&
  {Weaver}}{1994}]{woosley94}
{Woosley} S.~E.,  {Weaver} T.~A.,  1994, \mn@doi [\apj] {10.1086/173813}, \href
  {https://ui.adsabs.harvard.edu/abs/1994ApJ...423..371W} {423, 371}

\bibitem[\protect\citeauthoryear{{Worthey}, {Faber}  \& {Gonzalez}}{{Worthey}
  et~al.}{1992}]{Worthey1992}
{Worthey} G.,  {Faber} S.~M.,   {Gonzalez} J.~J.,  1992, \mn@doi [\apj]
  {10.1086/171836}, \href
  {https://ui.adsabs.harvard.edu/abs/1992ApJ...398...69W} {398, 69}

\bibitem[\protect\citeauthoryear{{Yanny} et~al.,}{{Yanny} et~al.}{2009}]{SEGUE}
{Yanny} B.,  et~al., 2009, \mn@doi [\aj] {10.1088/0004-6256/137/5/4377}, \href
  {https://ui.adsabs.harvard.edu/abs/2009AJ....137.4377Y} {137, 4377}

\bibitem[\protect\citeauthoryear{{de Boer}, {Belokurov}, {Beers}  \& {Lee}}{{de
  Boer} et~al.}{2014}]{deboer-sag-knee}
{de Boer} T.~J.~L.,  {Belokurov} V.,  {Beers} T.~C.,   {Lee} Y.~S.,  2014,
  \mn@doi [\mnras] {10.1093/mnras/stu1176}, \href
  {https://ui.adsabs.harvard.edu/abs/2014MNRAS.443..658D} {443, 658}

\bibitem[\protect\citeauthoryear{{de Jong} et~al.,}{{de Jong}
  et~al.}{2019}]{4most}
{de Jong} R.~S.,  et~al., 2019, \mn@doi [The Messenger]
  {10.18727/0722-6691/5117}, \href
  {https://ui.adsabs.harvard.edu/abs/2019Msngr.175....3D} {175, 3}

\bibitem[\protect\citeauthoryear{van~der Walt, Colbert  \& Varoquaux}{van~der
  Walt et~al.}{2011}]{vanderwalt_numpy}
van~der Walt S.,  Colbert S.~C.,   Varoquaux G.,  2011, \mn@doi [Computing in
  Science & Engineering] {10.1109/MCSE.2011.37}, 13, 22

\makeatother
\end{thebibliography}


\appendix


\bsp	
\label{lastpage}
\end{document}